\documentclass[useAMS,usenatbib]{mn2e}
\usepackage[dvips]{graphicx}
\usepackage{graphicx,times}
\usepackage{natbib}
\usepackage{rotating}
\usepackage[utf8]{inputenc}
\usepackage{amssymb,amsmath}
\usepackage[T1]{fontenc}
\usepackage{aecompl}
\bibpunct{(}{)}{;}{a}{}{,}
\title[Monthly Notices: \LaTeXe\ guide for authors]
  {Classification of 4XMM-DR9 Sources by Machine Learning}
\author[Y. Zhang et~al.]
  {Yanxia Zhang$^{1}$\thanks{Email: zyx@bao.ac.cn}
  , Yongheng Zhao$^{1}$, and Xue-Bing Wu$^{2,3}$ \\
  $^1$CAS Key Laboratory of Optical Astronomy, National Astronomical
Observatories, Beijing, 100101, China\\
  $^2$Department of Astronomy, School of Physics, Peking University, Beijing 100871, China.\\
  $^3$Kavli Institute for Astronomy and Astrophysics, Peking University, Beijing 100871, China.}
\date{Released 2002 Xxxxx XX}

\begin{document}

\label{firstpage}

\maketitle

\begin{abstract}
The ESA's X-ray Multi-Mirror Mission (\emph{XMM-Newton}) created a new, high quality version of the \emph{XMM-Newton} serendipitous source catalogue, 4XMM-DR9, which provides a wealth of information for observed sources. The 4XMM-DR9 catalogue is correlated with the Sloan Digital Sky Survey (SDSS) DR12 photometric database and the ALLWISE database, then we get the X-ray sources with information from X-ray, optical and/or infrared bands, and obtain the XMM-WISE sample, the XMM-SDSS sample and the XMM-WISE-SDSS sample. Based on the large spectroscopic surveys of SDSS and the Large Sky Area Multi-object Fiber Spectroscopic Telescope (LAMOST), we cross-match the XMM-WISE-SDSS sample with those sources of known spectral classes, and obtain the known samples of stars, galaxies and quasars. The distribution of stars, galaxies and quasars as well as all spectral classes of stars in 2-d parameter spaces is presented. Various machine learning methods are applied on different samples from different bands. The better classified results are retained. For the sample from X-ray band, rotation forest classifier performs the best. For the sample from X-ray and infrared bands, a random forest algorithm outperforms all other methods. For the samples from X-ray, optical and/or infrared bands, LogitBoost classifier shows its superiority. Thus, all X-ray sources in the 4XMM-DR9 catalogue with different input patterns are classified by their respective models which are created by these best methods. Their membership and membership probabilities to individual X-ray sources are assigned. The classified result will be of great value for the further research of X-ray sources in greater detail.

\end{abstract}


\section{Introduction} \label{sec:intro}

Since all X-rays are prevented from entering by the Earth's atmosphere, only a space-based telescope can observe and probe celestial X-ray sources. Both NASA's \emph{Chandra X-ray Observatory} and the ESA's X-ray Multi-Mirror Mission (\emph{XMM-Newton}) are space missions in X-ray band and further develop X-ray astronomy into a new era \citep{Bran05}. Significant discoveries have been found with these missions \citep{San09}. These missions may provide answers to other profound cosmic questions such as the enigmatic black holes, formation and evolution of galaxies, dark matter, dark energy, the origins of the Universe, and so on. They are taken as valuable tools to probe X-ray emission from various astrophysical systems. With the implementation of these missions, more and more X-ray sources are still not identified. Identification of deep X-ray survey sources is a challenging issue for several reasons \citep{Bran05}. Large sky survey projects (e.g. the Sloan Digital Sky Survey (SDSS), the Wide-field Infrared Survey Explorer (WISE), the Large Sky Area Multi-object Fiber Spectroscopic Telescope (LAMOST)) provide multiwavelength information and spectroscopic classes of X-ray sources. \citet{Pin11} cross-correlated the 2XMMi catalogue with SDSS DR7 and studied the high-energy properties of various classes of X-ray sources. Machine learning may learn knowledge from the known examples and create a classifier to predict unknown sources. Therefore machine learning makes it possible to classify X-ray sources depending on their multiwavelength and spectroscopic information of known samples. There are some works on this respect. For example, \citet{Bro11} applied a naive Bays classifier to classify X-ray sources from the $Chandra$ Carina Complex Project. \citet{Zhang13} performed random forest algorithm on the cross-matched sample between 2XMMi-DR3 and SDSS-DR8. \citet{Far15} classified the variable 3XMM Sources by the random forest algorithm. \citet{Arn20} identified new X-ray binary candidates in M31 also using the random forest algorithm.

In this paper, we download the 4XMM-DR9 catalogue, and obtain the spectroscopic classes of these X-ray sources from SDSS and LAMOST, X-ray information from \emph{XMM-Newton}, optical information from SDSS and infrared information from ALLWISE. We create the classifiers to classify the X-ray sources with known spectroscopic classes based on only X-ray information, combined X-ray and optical/infrared information, or combined X-ray, optical and infrared information. Section~2 describes the data used and the distribution of various objects in 2-d spaces. Section~3 presents the classification methodologies. Section~4 compares the performance of better classifiers for different samples. Section~5 discusses the results of the classifiers and applies the created classifiers to the unknown sources. Section~6 provides our conclusions for this work.

\section{The data} \label{sec:data}

The European Space Agency's (ESA) X-ray Multi-Mirror Mission (\emph{XMM-Newton}) was launched on December 10th 1999, performing in the X-ray, ultra-violet and optical bands. \emph{XMM-Newton} is ESA's second cornerstone of the Horizon 2000 Science Programme. It carries 3 high throughput X-ray telescopes with an unprecedented effective area, and an optical monitor, the first flown on an X-ray observatory. This mission had released a new and high quality version of the \emph{XMM-Newton} serendipitous source catalogue 4XMM-DR9. This catalogue includes 810 795 detections of 550 124 unique sources drawn from 11 204 \emph{XMM-Newton} EPIC observations, covering 1152 degrees$^2$ of the sky during the energy band from 0.2 keV to 12 keV \citep{Web20}. For the total photon energy band from 0.2 keV to 12 keV, the median flux of the catalogue detections is $\sim 2.3 \times 10^{-14}$ erg cm$^{-2}$ s$^{-1}$; it is $\sim 5.3 \times 10^{-15}$ erg cm$^{-2}$ s$^{-1}$ in the soft energy band (0.2-2 keV), and it is $\sim 1.2 \times 10^{-14}$ erg cm$^{-2}$ s$^{-1}$ in the hard band (2-12 keV). There are about 23 per cent of the sources with total fluxes below $1 \times 10^{-14}$ erg cm$^{-2}$ s$^{-1}$. The typical positional accuracy is about $2''$. For the astrometric quality, mean RA and Dec offsets between the XMM sources and the SDSS optical quasars are $-0.01''$ and $0.005''$ respectively with corresponding standard deviation of $0.70''$ and $0.64''$ (See fig.~10 in \citealt{Web20}).

The Wide-field Infrared Survey Explorer (WISE; \citealt{Wright10}) is an entire mid-infrared sky survey with simultaneous photometry in four filters at 3.4, 4.6, 12 and 22 $\mu$m ($W1, W2, W3$, and $W4$). It obtained over a million images and observed hundreds of millions of celestial objects. The WISE survey provides mid-infrared information about the Solar System, the Milky Way, and the Universe. On the basis of the WISE work, the AllWISE program has created new products with better photometric sensitivity and accuracy as well as better astrometric precision. The limiting magnitudes of $W1$ and $W2$ are brighter than 19.8 and 19.0 (Vega: 17.1, 15.7) for the ALLWISE source catalog. The sources brighter than 8, 7 in $W1$ and $W2$ bands are affected by saturation. Considering the accuracy of $W1, W2, W3$ and $W4$, we only adopt $W1$ and $W2$, converting $W1$ and $W2$ in Vega magnitudes to AB magnitudes by $W1_{AB}=W1+2.699$ and $W2_{AB}=W2+3.339$. The average PSF with FWHMs in $W1$, $W2$, $W3$ and $W4$ is $6.1''$, $6.4''$, $6.5''$ and $12''$, respectively. For high S/N ($>20$) sources, WISE positions are better than $0.15''$ for 1$\sigma$ and 1 axis.

The Sloan Digital Sky Survey (SDSS; \citealt{york00}) has been one of the most successful photometric and spectroscopic sky surveys ever made, providing deep multi-color images of one third of the sky and spectra for more than three million celestial objects. Data Release 12 (DR12) is the final data release of the SDSS-III, containing all SDSS observations through July 2014 \citep{Ei11}. It includes the complete dataset of the BOSS and APOGEE surveys, and also newly includes stellar radial velocity measurements from MARVELS. Data Release 16 (DR16) is the fourth SDSS data release (SDSS-IV; \citealt{Blan17}). SDSS mapped the sky in the five optical band passes ($ugriz$) with central wavelengths of 3551\AA, 4686\AA, 6165\AA, 7481\AA and 8931\AA. Pixel size is $0.396''$ and the astrometry accuracy is less than $0.1''$rms absolute per coordinate. The limiting magnitudes of $ugriz$ are 21.6, 22.2, 22.2, 21.3 and 20.7 at the 95 per cent completeness, respectively. For $u$ and $z$, they are converted to AB magnitudes by $u_{AB}=u-0.04$mag and $z_{AB}=z+0.02$mag. DR16 contains SDSS observations through August 2018, including 880 652 stars, 2 616 381 galaxies and 749 775 quasars when $zWarning=0$ in DR16 SpecObj database.

The Large Sky Area Multi-object Fiber Spectroscopic Telescope (LAMOST; \citealt{Cui12,Luo15}) may take 4000 spectra in a single exposure to a limiting magnitude as faint as $r=19$mag at the resolution $R=1800$. It has finished the first five year survey plan. LAMOST survey contains the LAMOST ExtraGAlactic Survey (LEGAS), and the LAMOST Experiment for Galactic Understanding and Exploration (LEGUE) survey of Milky Way stellar structure. The data products of the fifth data release (DR5) include 8 183 160 stars (7 540 605 stars with S/N in $g$ band or $i$ band larger than 10), 152 863 galaxies, 52 453 quasars, and 637 889 unknown objects.

The SDSS Data Release 14 Quasar Catalogue (DR14Q;
\citealt{Paris18}) contains 526 356
spectroscopically identified quasars. DR14Q consists of spectroscopically identified quasars from SDSS-I, II, III and the latest SDSS-IV eBOSS survey.

In order to obtain multiwavelength properties of X-ray sources, we cross-match the 4XMM-DR9 catalog with SDSS and ALLWISE databases. According to the work of \citep{Cov08}, we estimate spurious SDSS and ALLWISE matches by applying a 30$''$ offset to the X-ray source declinations and searching the 4XMM-DR9 catalog for the sources with SDSS counterparts within $8''$ and ALLWISE counterparts within $10''$ for each X-ray source centroid. Figure~1 shows the normalized cumulative histogram of separation between 4XMM-DR9 and SDSS sources as well as 4XMM-DR9 and ALLWISE sources; the solid histogram represents the cumulative distribution of separation between X-ray and optical counterparts for real XMM-SDSS sources within $8''$ (left panel of Figure~1) and that of separation between X-ray and infrared counterparts for real XMM-WISE sources within $10''$ (right panel of Figure~1); the dashed histogram indicates the upper limit to the fractional contamination of the XMM-SDSS sample by chance superpositions of independent X-ray and optical sources (left panel of Figure~1) and the XMM-WISE sample by chance superpositions of independent X-ray and infrared sources (right panel of Figure~1). In general, high completeness and low contamination can not be achieved at the same time, higher completeness is needed at the expense of contamination, otherwise if we pursue low contamination, we must sacrifice completeness. For XMM matching SDSS at 3$''$, 4$''$, 5$''$ and 6$''$, the completeness vs. contamination is respectively 71.68 per~cent vs. 9.75 per~cent, 79.49 per~cent vs. 16.53 per~cent, 85.55 per~cent vs. 24.50 per~cent, 90.78 per~cent vs. 33.36 per~cent; for XMM matching ALLWISE at 3$''$, 6$''$, 7$''$ and 8$''$, the completeness vs. contamination is respectively 56.90 per~cent vs. 0.89 per~cent, 77.66 per~cent vs. 1.75 per~cent, 83.39 per~cent vs. 2.01 per~cent, 89.14 per~cent vs. 2.08 per~cent. From Figure~1, the fraction of X-ray sources matching SDSS occupies over 90 per~cent at $6''$ and that matching ALLWISE is about 90 per~cent at $8''$. So the cross-match radius between SDSS and 4XMM-DR9 sources is set as $6''$ while that between ALLWSIE and 4XMM-DR9 sources is adopted as $8''$. We apply the software TOPCAT \citep{Tay05} to perform cross-match. Finally we obtain the XMM-WISE sample and the XMM-SDSS sample, then the XMM-WISE-SDSS sample is derived according to the same ID (srcid) in the XMM-WISE sample and XMM-SDSS sample. All photometries throughout this paper are extinction-corrected according to the work \citep{Sch17} and AB magnitudes are adopted.

\begin{figure*}
\centering
\includegraphics[bb=1 1 430 369,width=8cm]{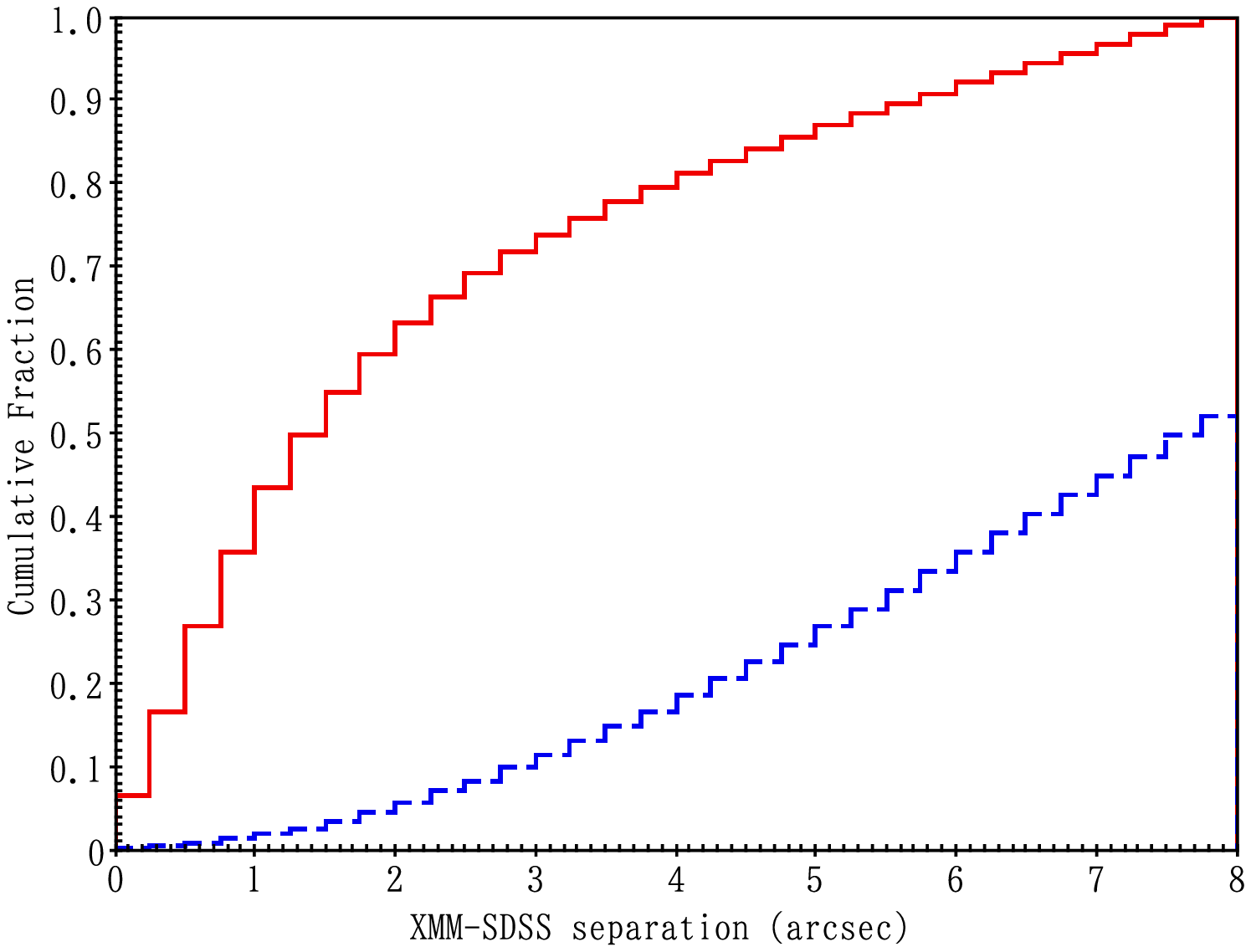}
\includegraphics[bb=1 1 430 369,width=7.8cm]{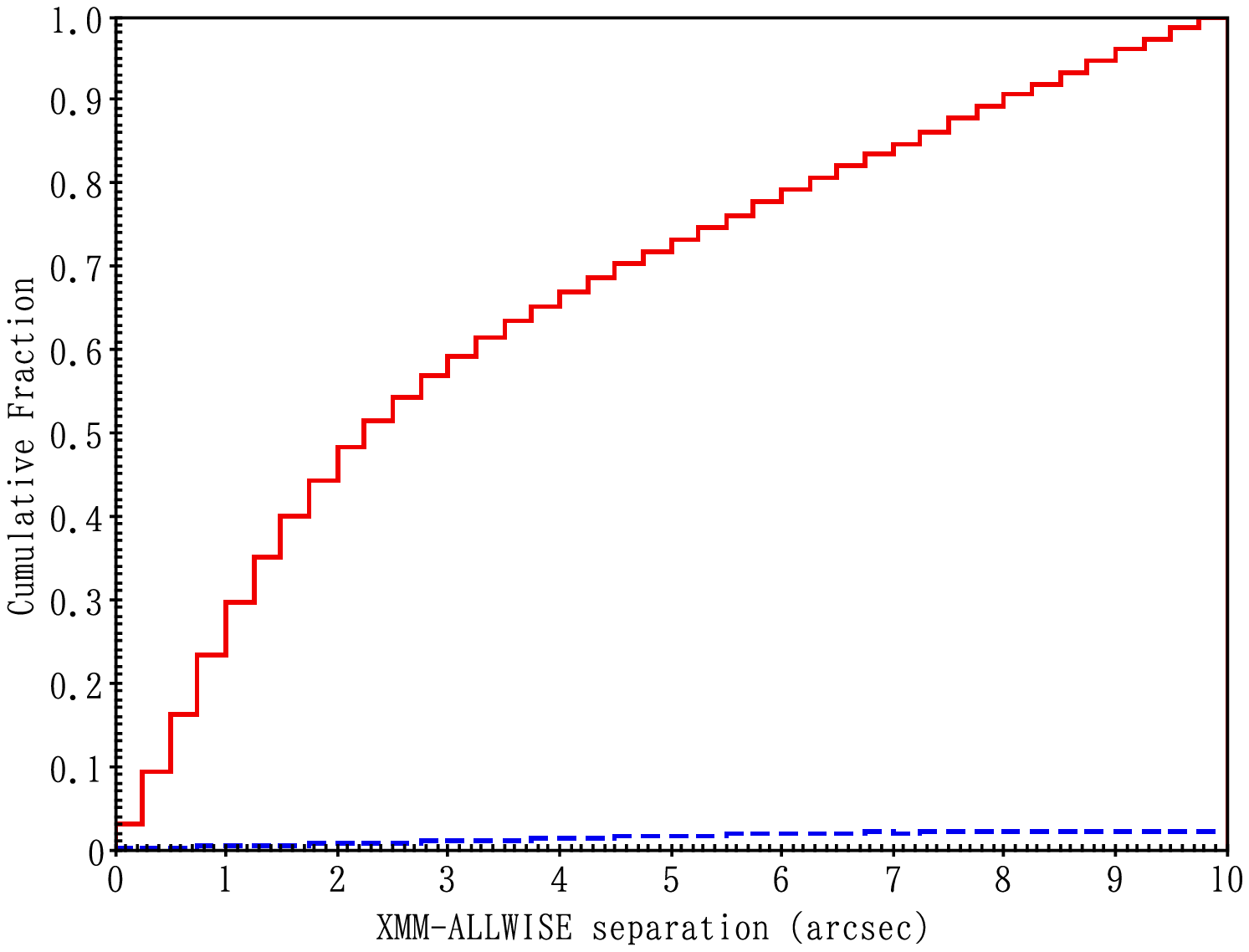}
\caption{Histogram of sources separation between XMM and SDSS as well as between XMM and ALLWISE. \textit{Red solid histogram}: cumulative distribution of separation between X-ray and optical counterparts (left panel) and between X-ray and infrared counterparts (right panel) for the real X-ray sources; \textit{blue dashed line}: distribution of separations returned by matching the faked X-ray sources with coordinates shifted by 30 arcsec to SDSS database (left panel) and ALLWISE database (right panel).}
\label{fig1}
\end{figure*}

In order to construct the known spectral samples, the samples have been identified spectroscopically by SDSS DR16 and LAMOST DR5. The known samples are cross-matched with the XMM-WISE-SDSS sample in 6 arcsec radius. Keeping the data quality, $zWarning=0$ is set in DR16 SpecObj database when downloading data, sc\_poserr$\le5$ and sc\_sum\_flag$<3$ are set in the 4XMM-DR9 database, the records with default values of $ugriz$, $W1$ and $W2$ are removed, the records with $W1<8$ and $W2<7$ are deleted, stars in LAMOST DR5 database are adopted with S/N in $g$ band or $i$ band larger than 10. When the objects are both identified by SDSS and LAMOST, the spectral class of objects in SDSS are only maintained. If the objects with known spectral class in the XMM-WISE-SDSS sample have counterparts in DR14Q, the objects are labelled as QSO. Finally, the known samples contain 3 558 stars, 7 203 galaxies and 21 040 quasars with information from X-ray, optical and infrared bands. The spectra were identified as STAR, GALAXY and QSO by SDSS and LAMOST automated classification pipeline using template fitting. The detailed information about known samples are indicated in Table~1. For the class assigned as GALAXY, the subclass as Non is from the LAMOST database and the subclass as default value is from the SDSS database. The LAMOST pipeline doesn't provide the subclasses for galaxies, and all the subclasses of galaxies in LAMOST database are labelled as Non. The websites related to the above datasets are shown in Table~2. About definitions and abbreviations in Table~1, AGN is short for active galactic nuclei, AGN BL for BROADLINE AGN, SB for starburst galaxy, SB BL for BROADLINE SB, SF for star forming galaxy, SF BL for BROADLINE SF, BL for BL Lacertae objects, CV for cataclysmic variable star, EM for emission line star, WD for white dwarf, DB for double or binary star, sdM1 for subdwarf M1 star, Carbon for carbon star, O, B, A, F, G, K, M for stars with spectral types of O, B, A, F, G, K, M, respectively. All these subclasses are assigned by SDSS and LAMOST automated classification pipeline depending on the spectroscopic characteristics. \citet{Blan12} presented that the galaxy spectra from SDSS by the line-fitting code were grouped into AGN, SF, and SB, if the spectra meet log10([OIII]/H$\beta$)$>$1.2 log10([NII]/H$\alpha$)$+0.22$, the galaxy spectra were identified as AGN, otherwise for the equivalent width (EW) of H$\alpha$: SF if EW(H$\alpha$)$<$50\AA, and SB if EW(H$\alpha$)$>$50\AA; galaxies and quasars may be classified as BROADLINE (BL) when their line widths are larger than 200 km s$^{-1}$; stellar spectra were classified as spectral types from O to M based on the ELODIE stellar library. The BROADLINE classification given by the SDSS pipeline does not necessarily indicate that an AGN has emission lines broad enough to classify it as a broad line (as opposed to narrow line) AGN because the emission line widths are typically more than 2000 km s$^{-1}$ for broad line AGNs \citep{Hao05}. BL Lacertae objects are a subclass of AGNs, which have fast and large amplitude variability over the whole spectra, high and variable polarization, and continuous spectra with no or weak absorption and emission features. Starburst galaxies are characterized by higher rates of star formation than normal galaxies. They are either young or rejuvenated galaxies, which typically contain very luminous X-ray sources. Since the separation of subclasses of galaxies depends on spectral line information, it seems difficult to discriminate them without spectra.

\begin{table}
\begin{center}
\caption[]{The number of class and subclass for the known samples\label{tab:Sample}}
 \begin{tabular}{rll}
 \hline
 \hline
 Class   &Subclass&No.\\
 \hline
 GALAXY  &AGN     &611  \\
         &AGN BL  &107  \\
         &SB      &387  \\
         &SB BL   &8    \\
         &SF      &1 008 \\
         &SF BL   &46   \\
         &BL      &281  \\
         &Non     &219  \\
         &        &4 536 \\
 \hline
 STAR    &O       &1    \\
         &B       &5    \\
         &A       &79   \\
         &F       &708  \\
         &G       &869  \\
         &K       &777  \\
         &M       &1 062 \\
         &CV      &39   \\
         &DB      &5   \\
         &EM      &1    \\
         &WD      &10   \\
         &sdM1    &1    \\
         &Carbon  &1    \\
 \hline
 QSO     &        &21 040\\
 \hline
\end{tabular}
\end{center}
\end{table}

\begin{table}
\begin{center}
\caption[]{The website for related catalogs \label{tab:Web}}
 \begin{tabular}{l}
 \hline
 \hline
4XMM-DR9 catalogue\\
https://www.cosmos.esa.int/web/xmm-newton/xsa\\
\hline
Spectrally identified stars, galaxies and quasars from SDSS\\
http://skyserver.sdss.org/dr16/en/tools/search/sql.aspx\\
\hline
Spectrally identified stars, galaxies and quasars from LAMOST\\
http://dr5.lamost.org/v3/catalogue\\
\hline
SDSS DR14 Quasar catalog (DR14Q)\\
https://www.sdss.org/dr14/algorithms/qso\_catalog\\
\hline
\end{tabular}
\end{center}
\end{table}

We select the features [log$(f_x),hr1,hr2,hr3,hr4,extent,r,\\W1,u-g,g-r,r-i,i-z,z-W1,W1-W2$,log$(f_x/f_r)$] of the known tar, galaxy and quasar samples used for this study. The selected features are described in Table~3. The 2-d plots between two attributes from these features are described in Figures~2 and 3. Figure~2 shows the difference among stars, galaxies and quasars, while Figure~3 indicates the difference among different spectral classes of stars. The two figures tell us that it is difficult to discriminate stars, galaxies and quasars, and different spectral classes of stars depending only on two attributes. These attributes all contribute more or less to the classification. As shown in Figure~2, most quasars obviously have larger log$(f_x/f_r)$, $r$, $W1$ and $W1-W2$ values than stars. It is easy to classify stars and quasars from galaxies with the attribute $extent$ in X-ray band. Nevertheless some AGNs do not appear as X-ray extended if emission is nuclear-dominated, thus they are misclassified as stars or quasars only depending on $extent$. We check stars and quasars with large $extent$ in SIMBAD and NED within 3$''$ radius, and find that some of them are galaxy in group of galaxies, galaxy cluster, or other kinds of objects. Most of galaxies indeed have relatively larger $extent$ in X-ray band. Most galaxies overlap most quasars with X-ray and infrared information while most galaxies overlap most stars with X-ray, optical and infrared information. In order to effectively separate stars, galaxies and quasars, it is necessary to apply all available information. As indicated in Figure~3, CV stars have more strong X-ray emission than other stars, most CV stars and M stars have more strong infrared emission than the remainder of the stars. They can be separated easily from the star sample in some 2-d spaces. Apparently they have obvious differences from the remainder of the star sample as they are mixed together and thus difficult to discriminate.

\begin{table*}
\begin{center}
\caption{The parameters, definition, catalogues and wavebands
}
\begin{tabular}{rlll}
\hline \hline
Parameter&Definition  &Catalogue& Waveband\\
\hline
srcid &source ID      &\emph{XMM}&X-ray band\\
sc\_ra &Right ascension in decimal degrees&\emph{XMM}&X-ray band\\
sc\_dec&Declination in decimal degrees    &\emph{XMM}&X-ray band\\
$hr1$&Hardness ratio 1 &\emph{XMM}&X-ray band\\
    &Definition: $hr1=(B-A)/(B+A)$, where&&\\
    &A=countrate in energy band 0.2-0.5keV&&\\
    &B=countrate in energy band 0.5-1keV&&\\
$hr2$&Hardness ratio 2 &\emph{XMM}&X-ray band\\
    &Definition: $hr2=(C-B)/(C+B)$, where&&\\
    &B=countrate in energy band 0.5-1keV&&\\
    &C=countrate in energy band 1-2keV&&\\
$hr3$&Hardness ratio 3 &\emph{XMM}&X-ray band\\
    &Definition: $hr3=(D-C)/(D+C)$, where&&\\
    &C=countrate in energy band 1-2keV&&\\
    &D=countrate in energy band 2-4.5keV&&\\
$hr4$&hardness ratio 4 &\emph{XMM}&X-ray band\\
    &Definition: $hr4=(E-D)/(E+D)$, where&&\\
    &D=countrate in energy band 2-4.5keV&&\\
    &E=countrate in energy band 4.5-12keV&&\\
extent&Source extent&\emph{XMM}&X-ray band\\
log$(f_{\rm x})$&X-ray flux&\emph{XMM}&X-ray band\\
log$(f_{\rm x}/f_{\rm r})$&X-ray-to-optical flux ratio&SDSS,\emph{XMM}&Optical and X-ray bands\\
$u$  & $u$ magnitude&SDSS& Optical band\\
$g$  & $g$ magnitude&SDSS& Optical band\\
$r$  & $r$ magnitude&SDSS& Optical band\\
$i$  & $i$ magnitude&SDSS& Optical band\\
$z$  & $z$ magnitude&SDSS& Optical band\\
$W1$  & $W1$ magnitude&ALLWISE& Infrared band\\
$W2$  & $W2$ magnitude&ALLWISE& Infrared band\\
\hline
\end{tabular}
\bigskip
\end{center}
\end{table*}

\begin{figure*}
\centering
\includegraphics[bb=1 1 427 356,width=5cm]{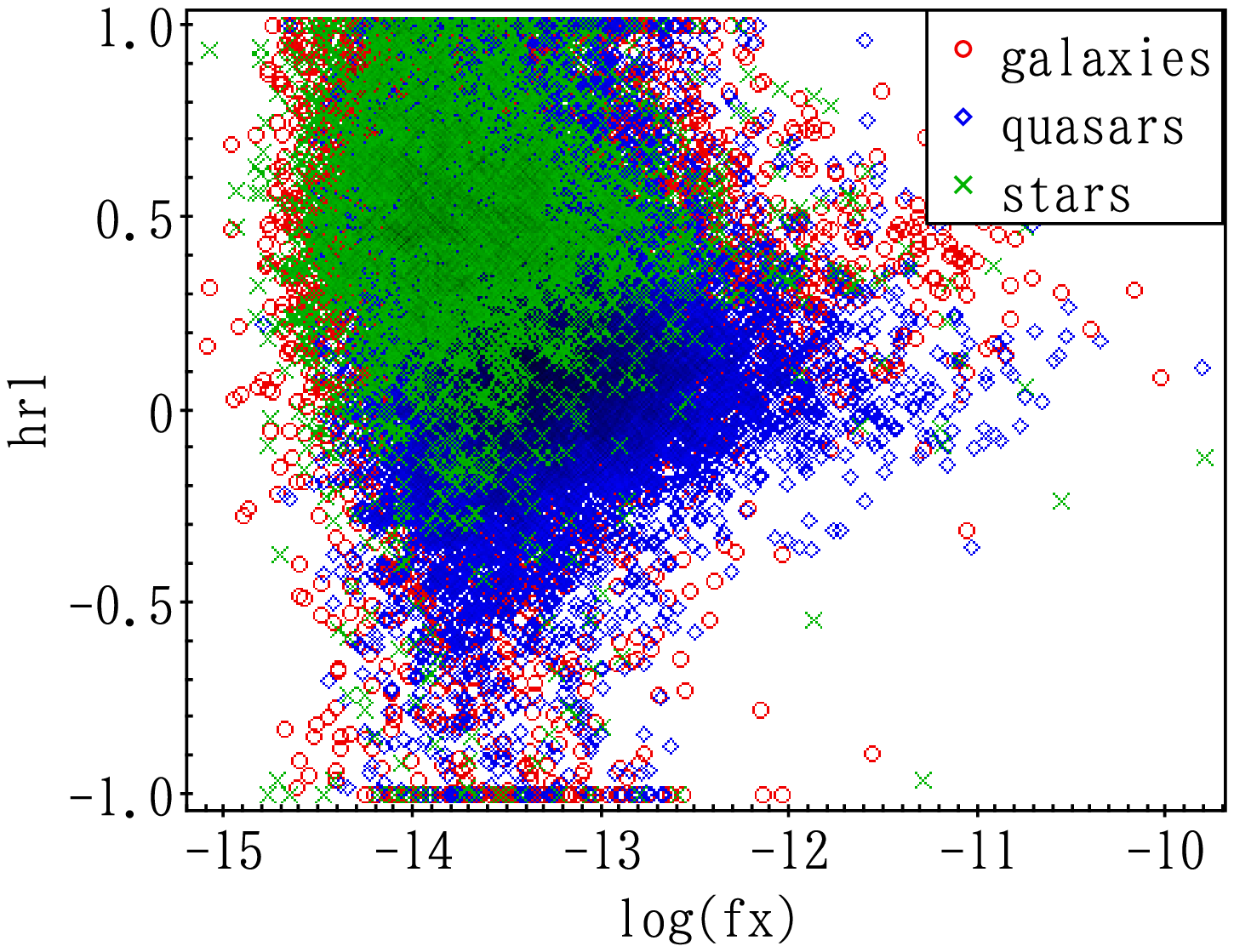}
\includegraphics[bb=1 1 427 356,width=5cm]{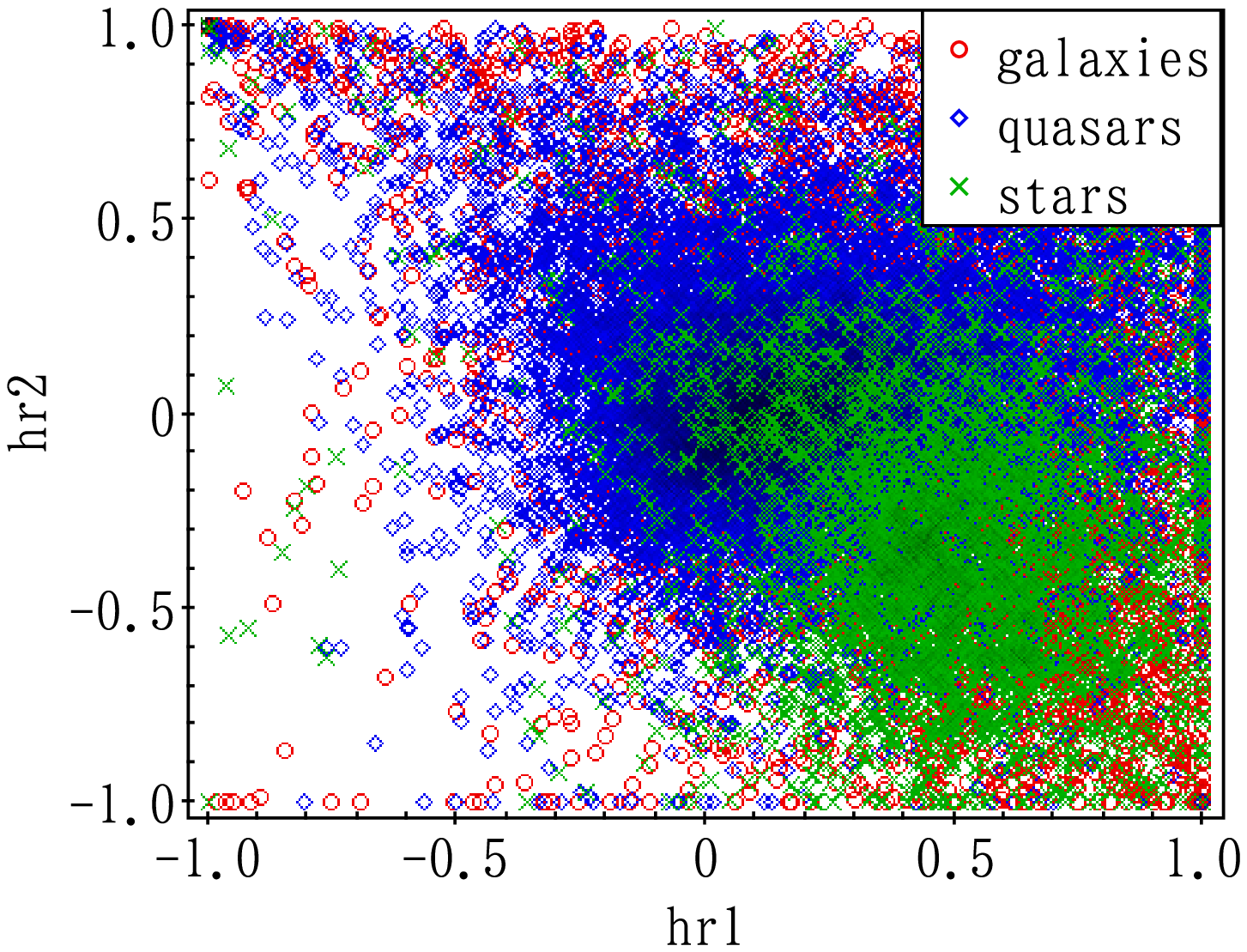}
\includegraphics[bb=1 1 427 356,width=5cm]{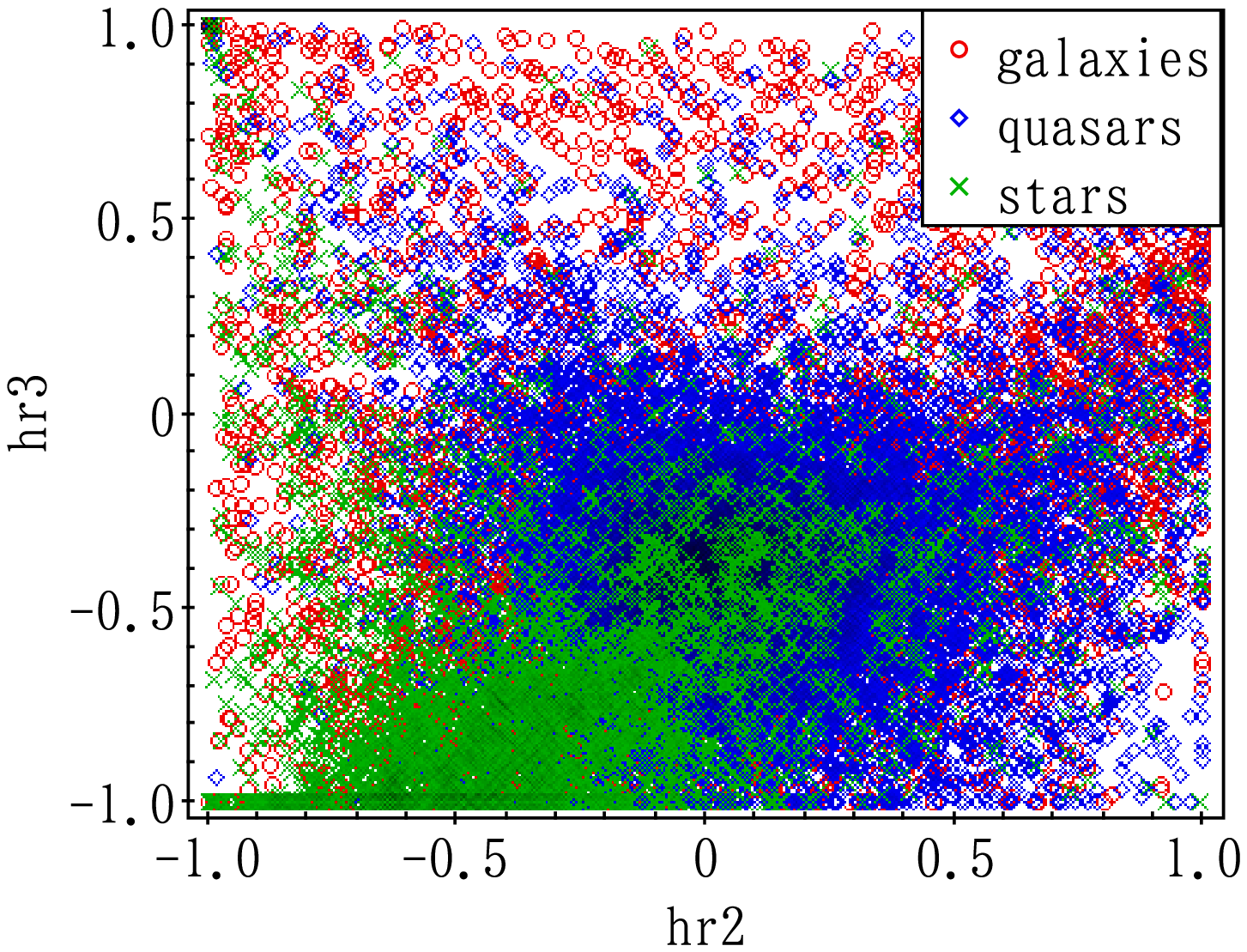}
\includegraphics[bb=1 1 427 356,width=5cm]{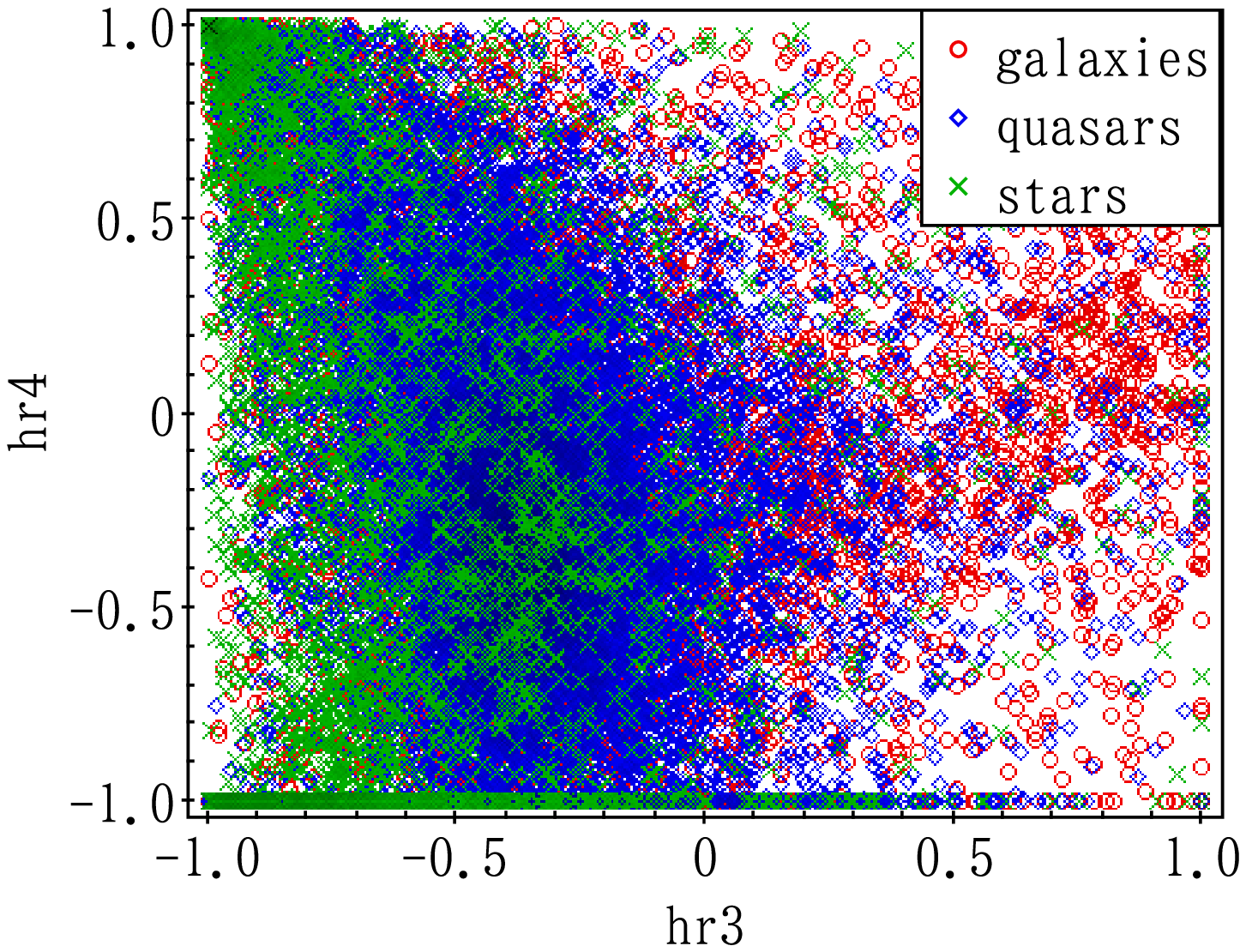}
\includegraphics[bb=1 1 427 356,width=5cm]{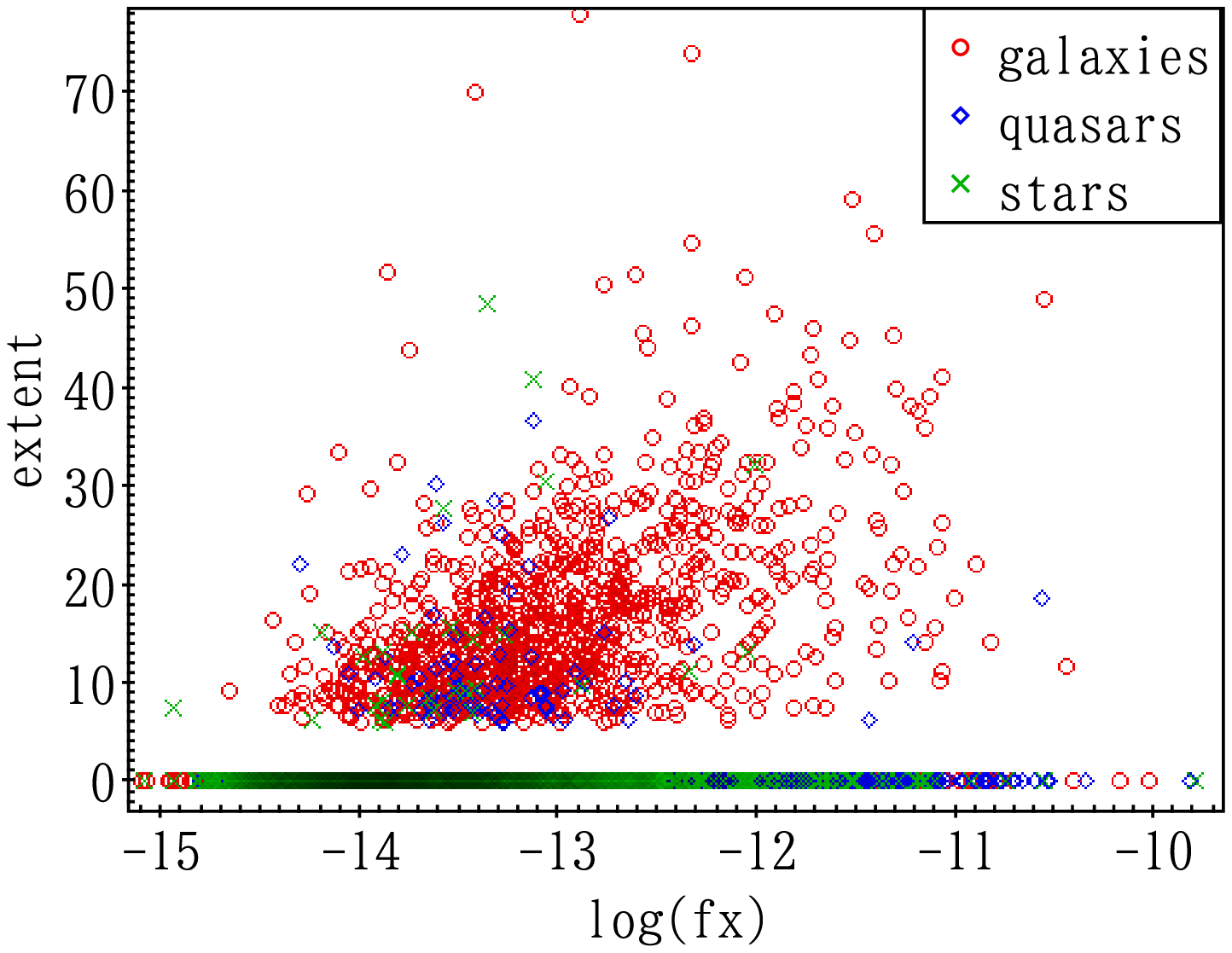}
\includegraphics[bb=1 1 427 356,width=5cm]{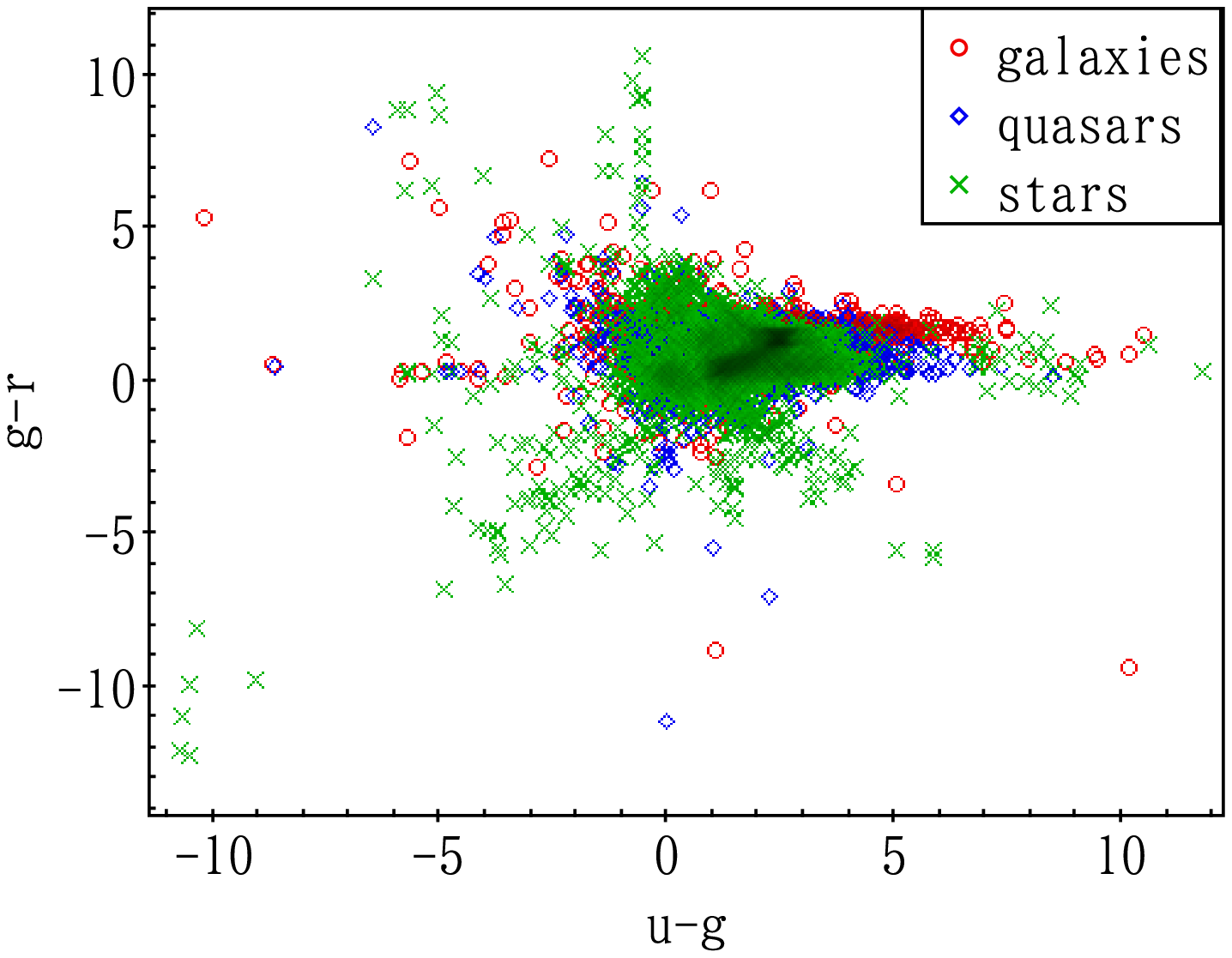}
\includegraphics[bb=1 1 427 356,width=5cm]{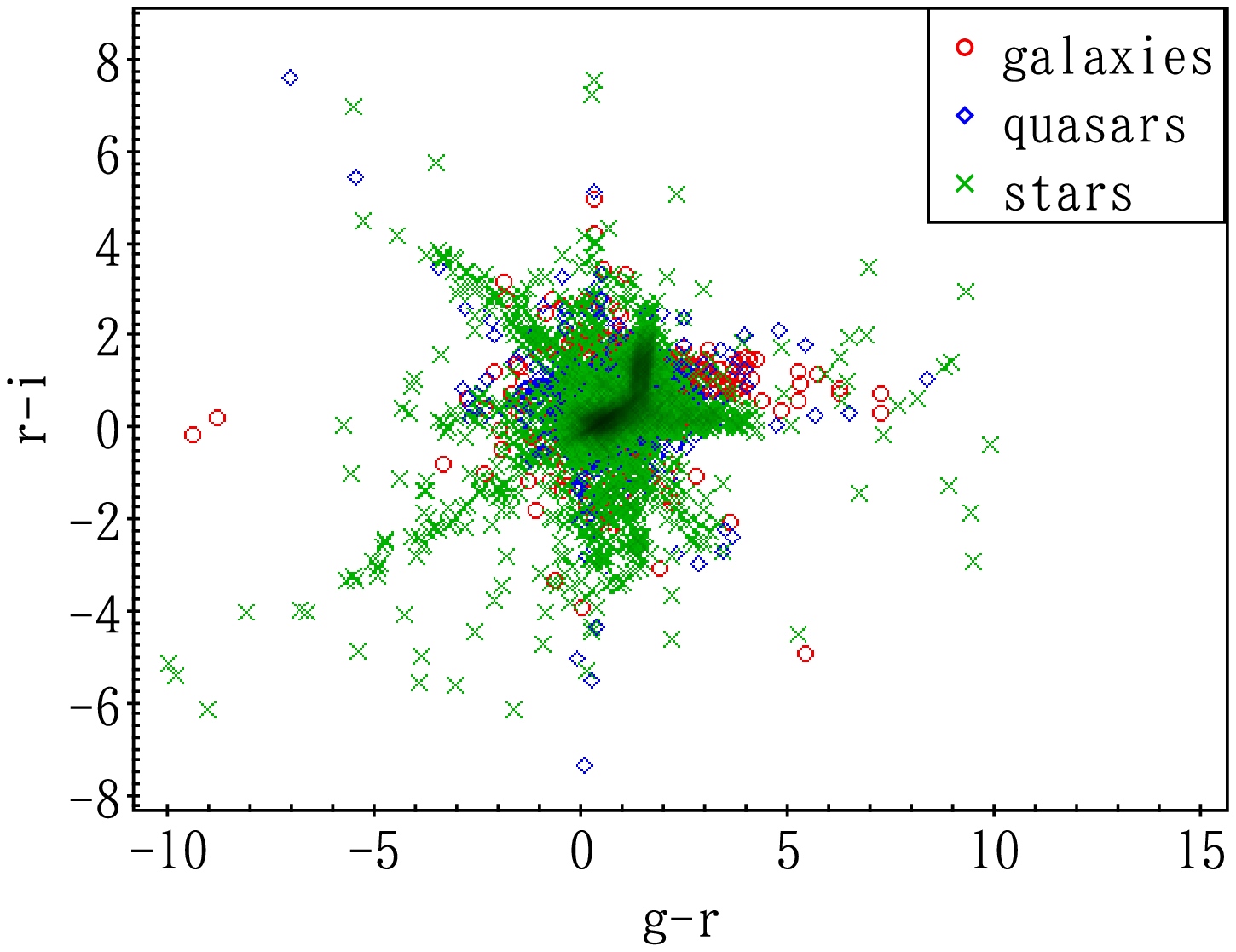}
\includegraphics[bb=1 1 427 356,width=5cm]{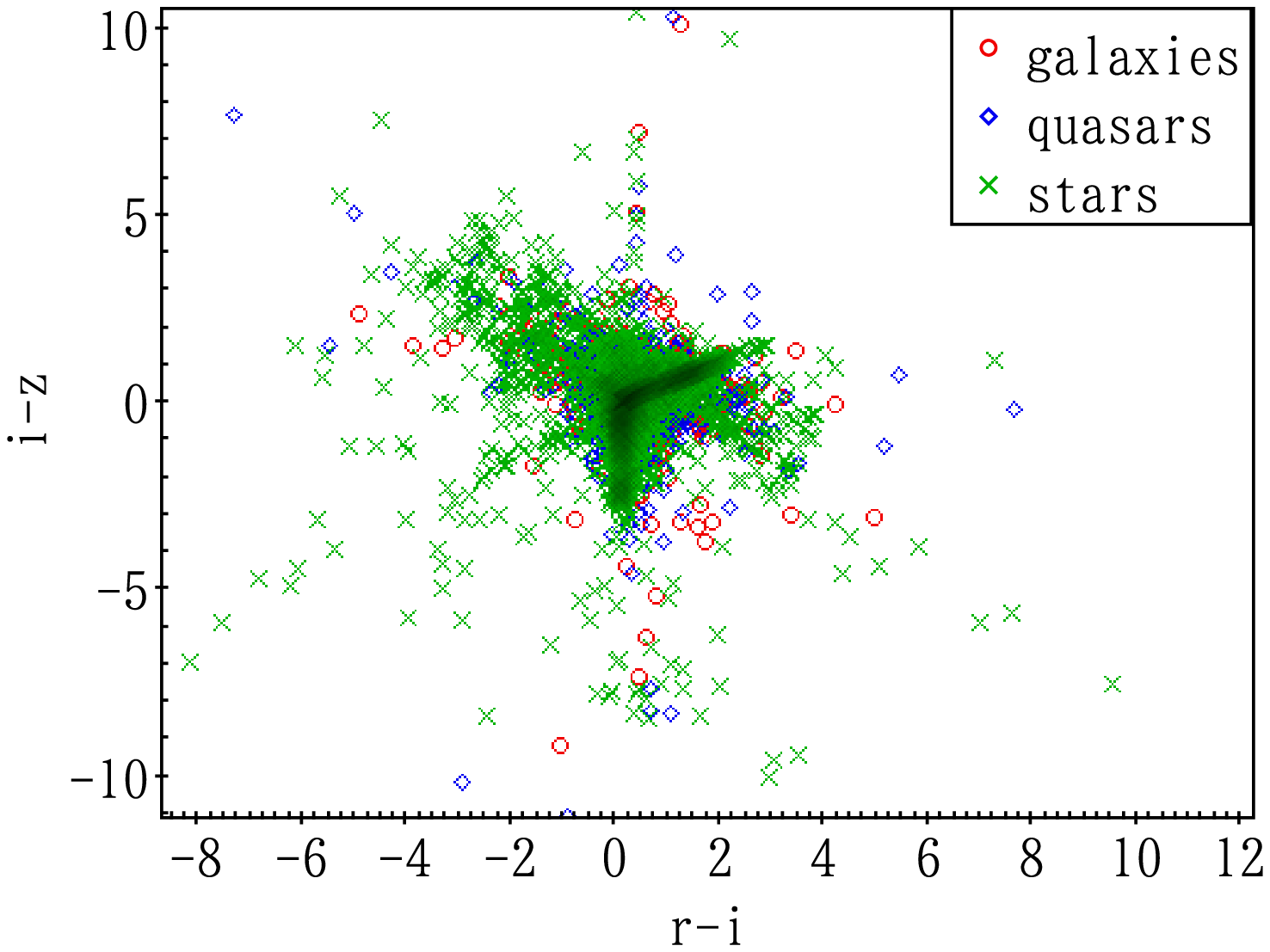}
\includegraphics[bb=1 1 427 356,width=5cm]{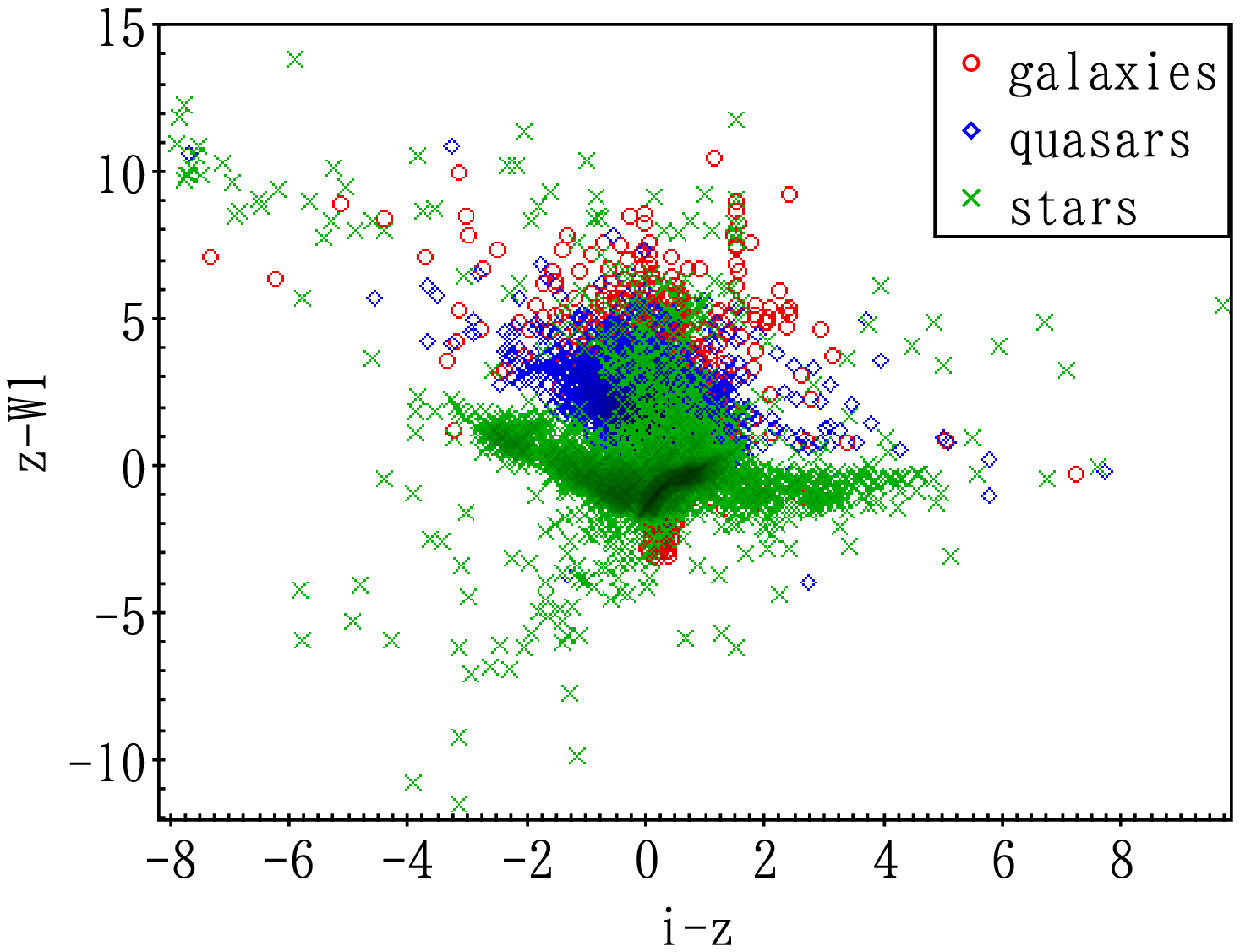}
\includegraphics[bb=1 1 427 356,width=5cm]{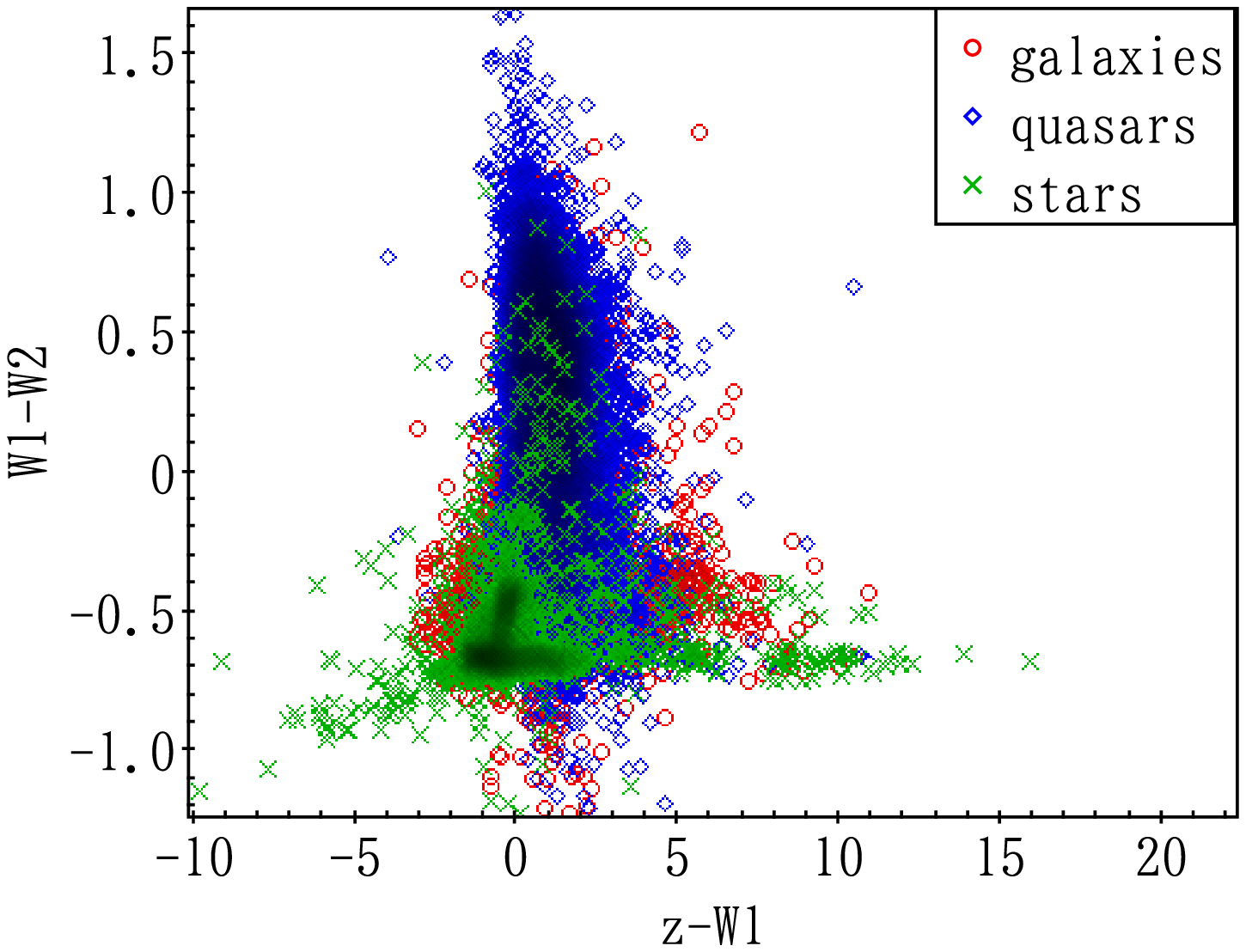}
\includegraphics[bb=1 1 427 356,width=5cm]{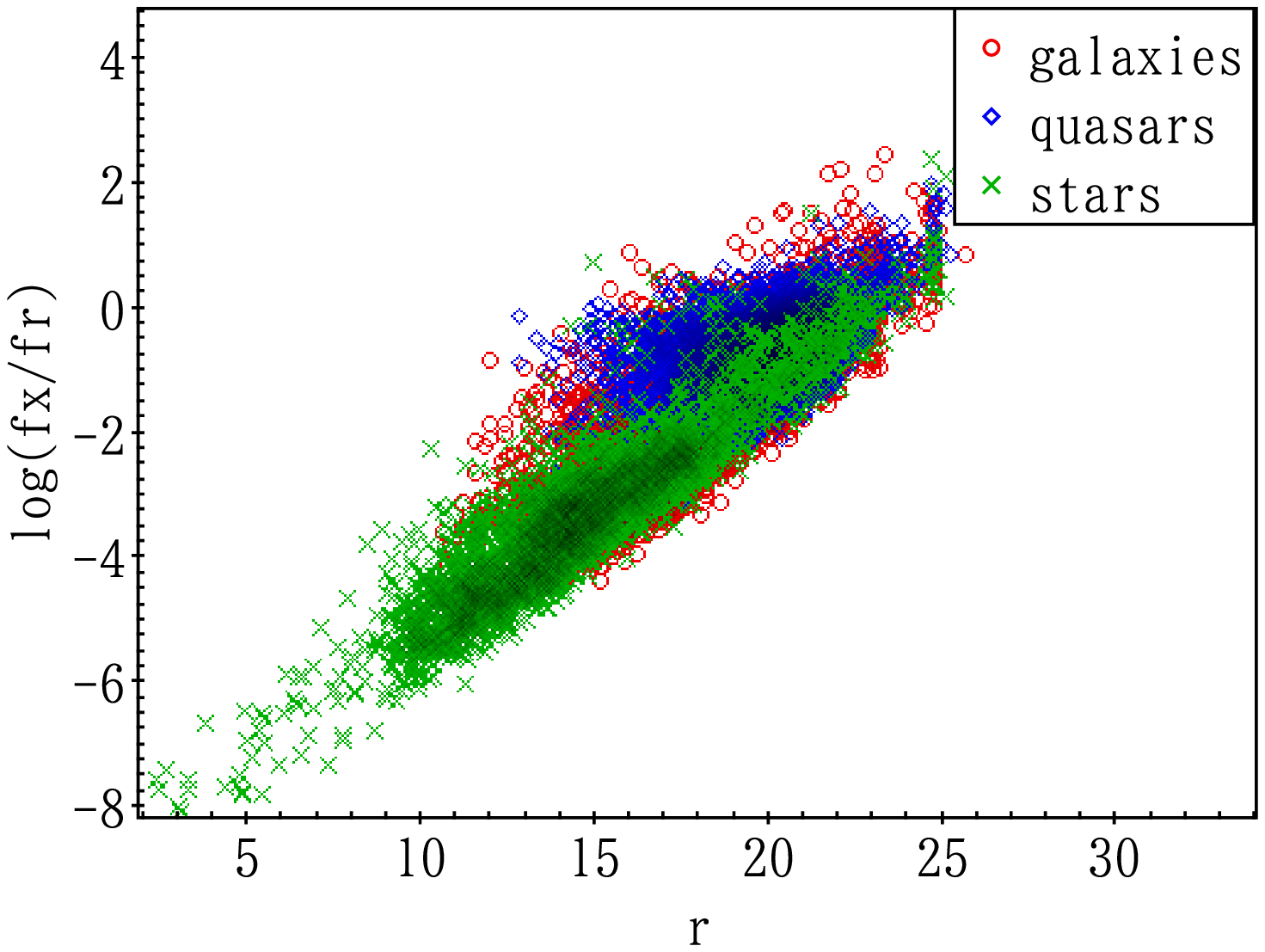}
\includegraphics[bb=1 1 427 356,width=5cm]{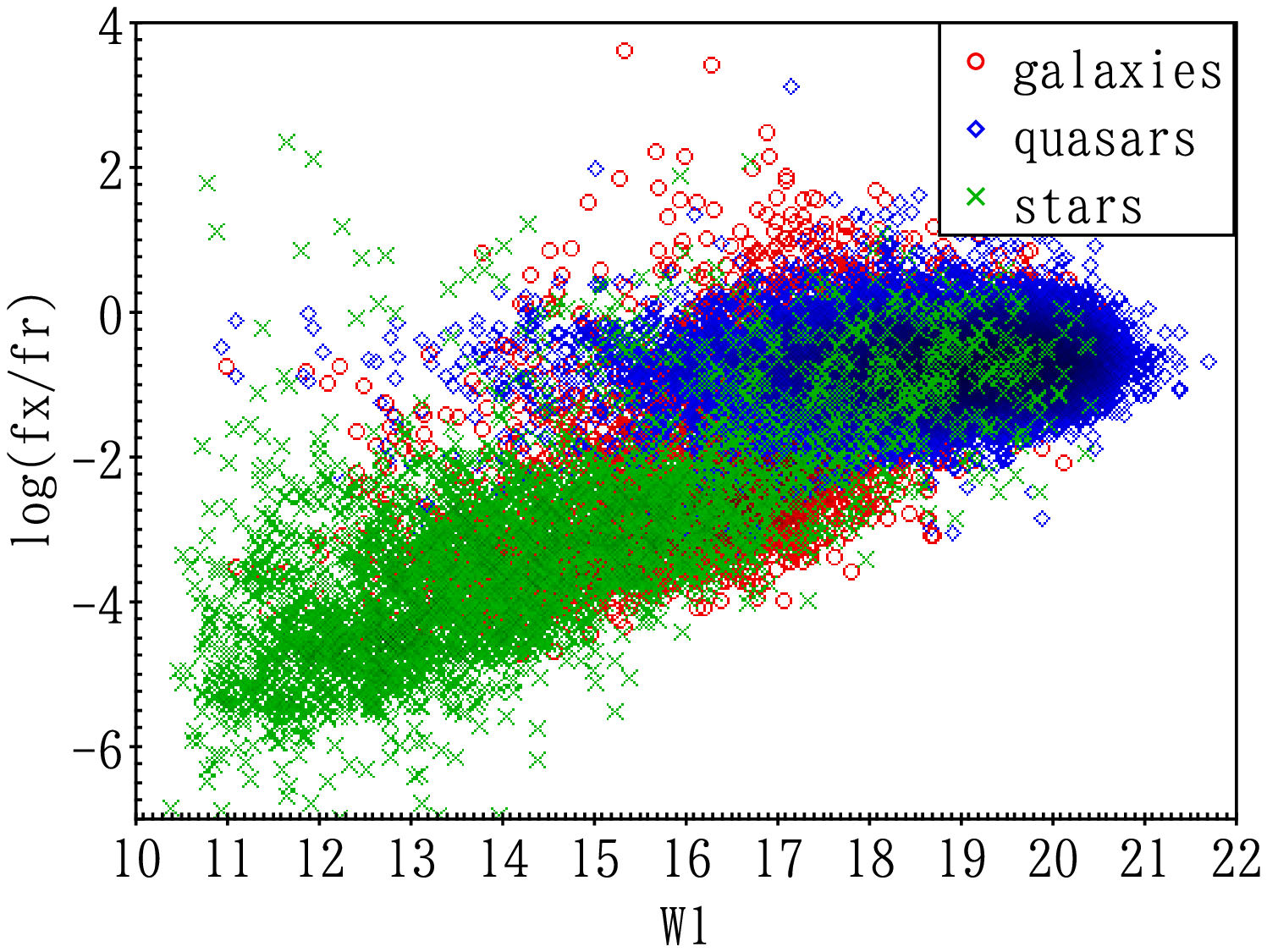}
\includegraphics[bb=1 1 427 356,width=5cm]{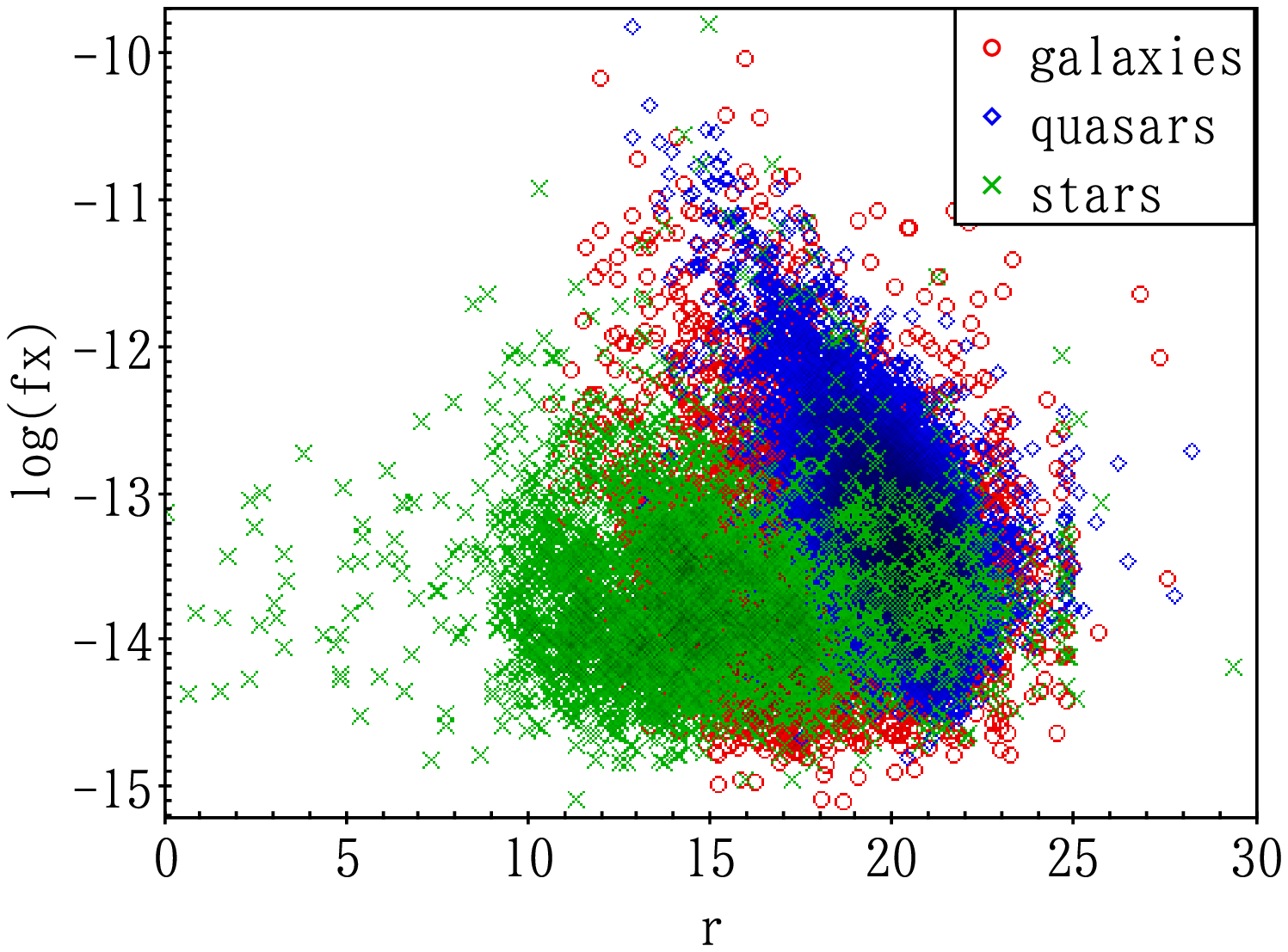}
\includegraphics[bb=1 1 427 356,width=5cm]{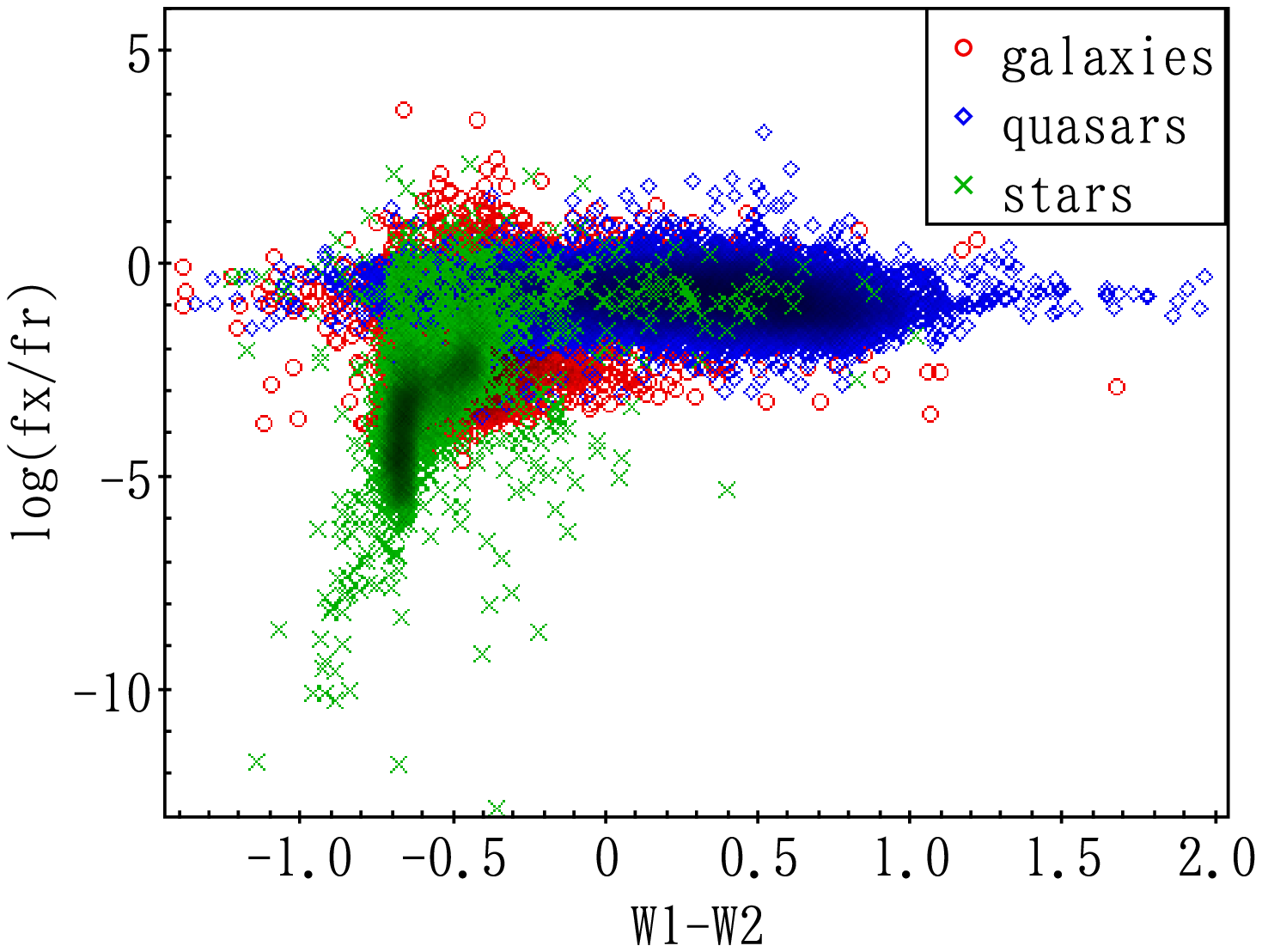}
\includegraphics[bb=1 1 427 356,width=5cm]{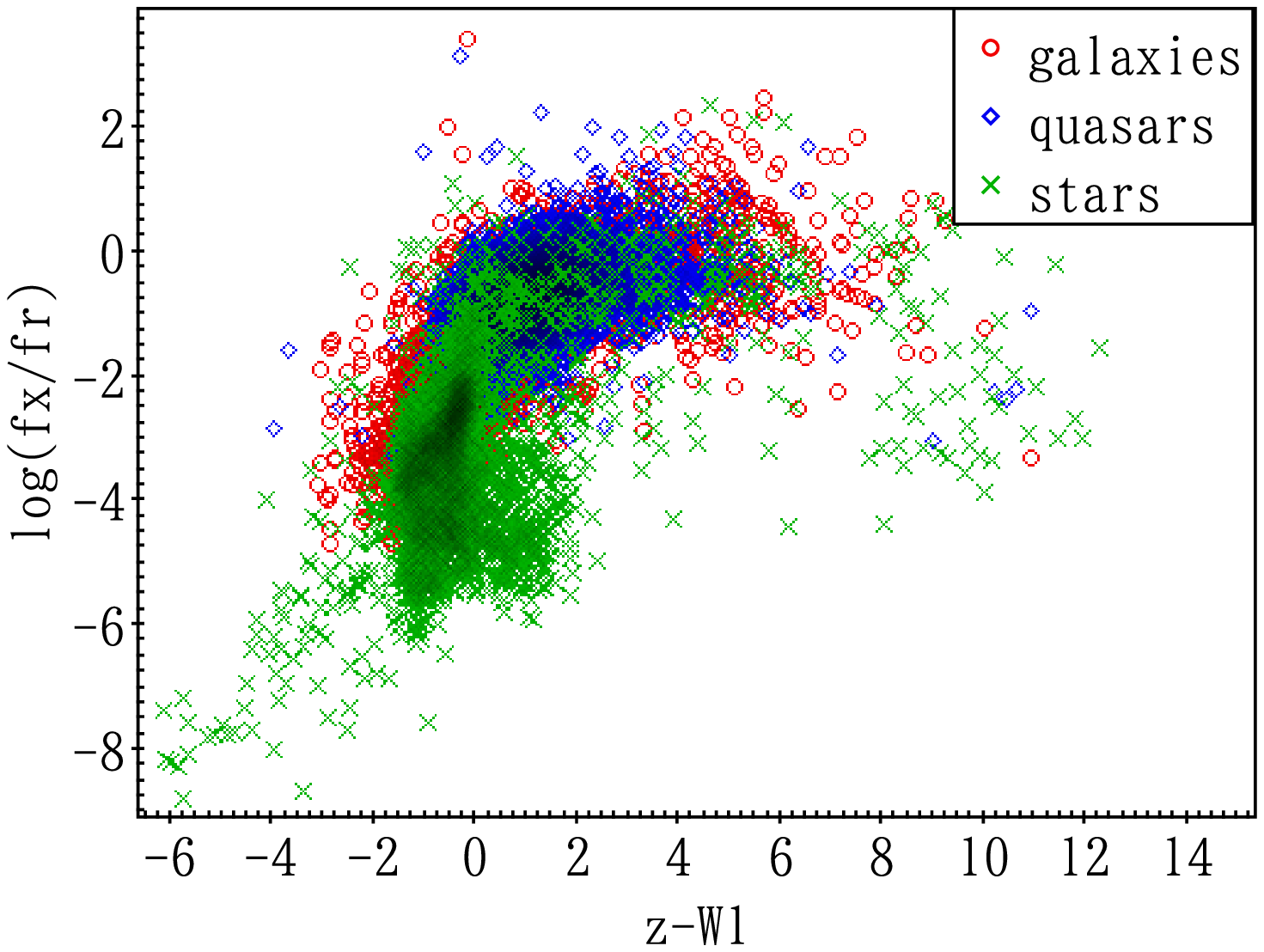}
\caption{The distribution of stars, galaxies and quasars in 2-d spaces, red open circles represent galaxies, blue open diamonds represent quasars and green crosses represent stars.}
\label{fig2}
\end{figure*}

\begin{figure*}
\centering
\includegraphics[bb=-1 -1 427 356,width=5cm]{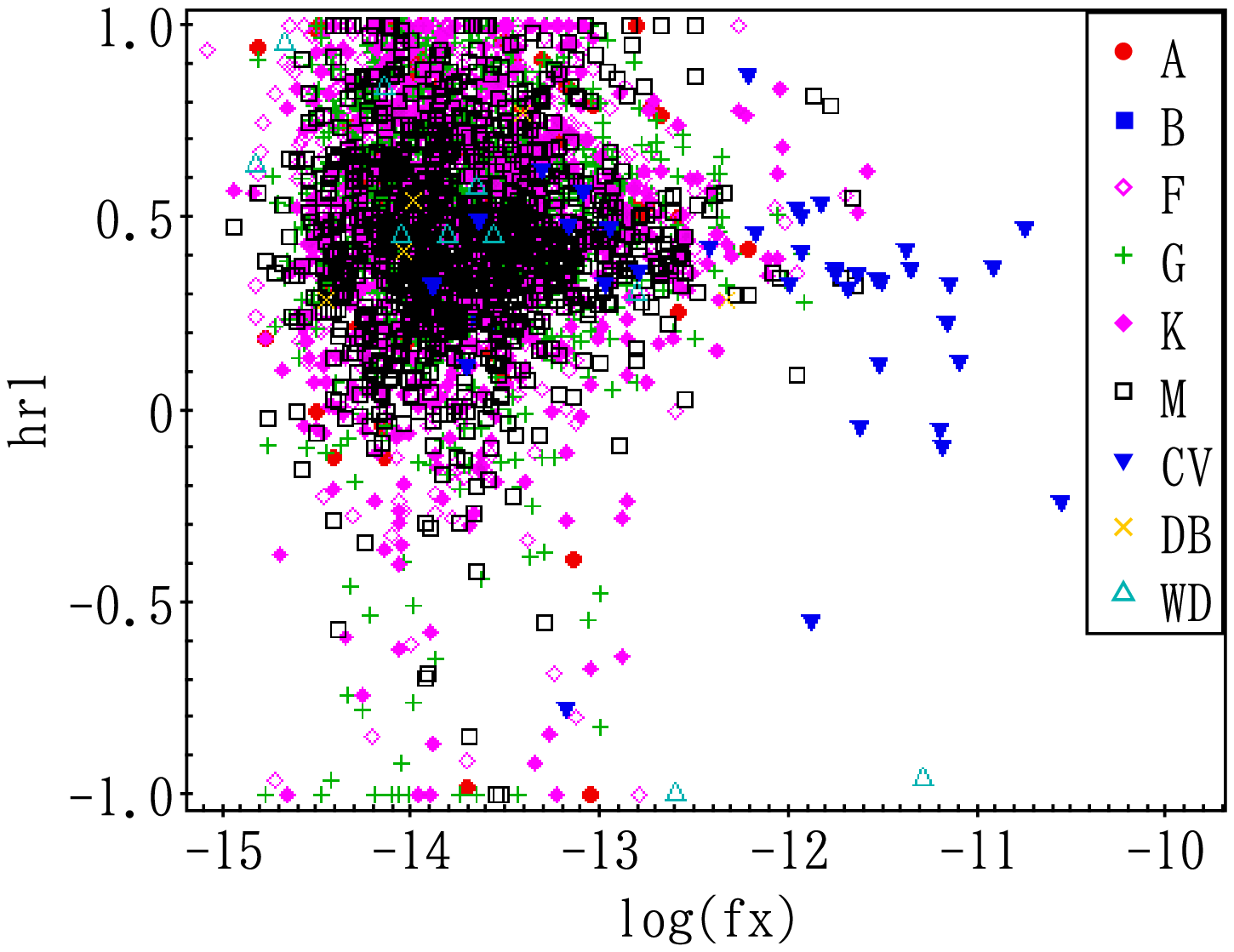}
\includegraphics[bb=-1 -1 427 356,width=5cm]{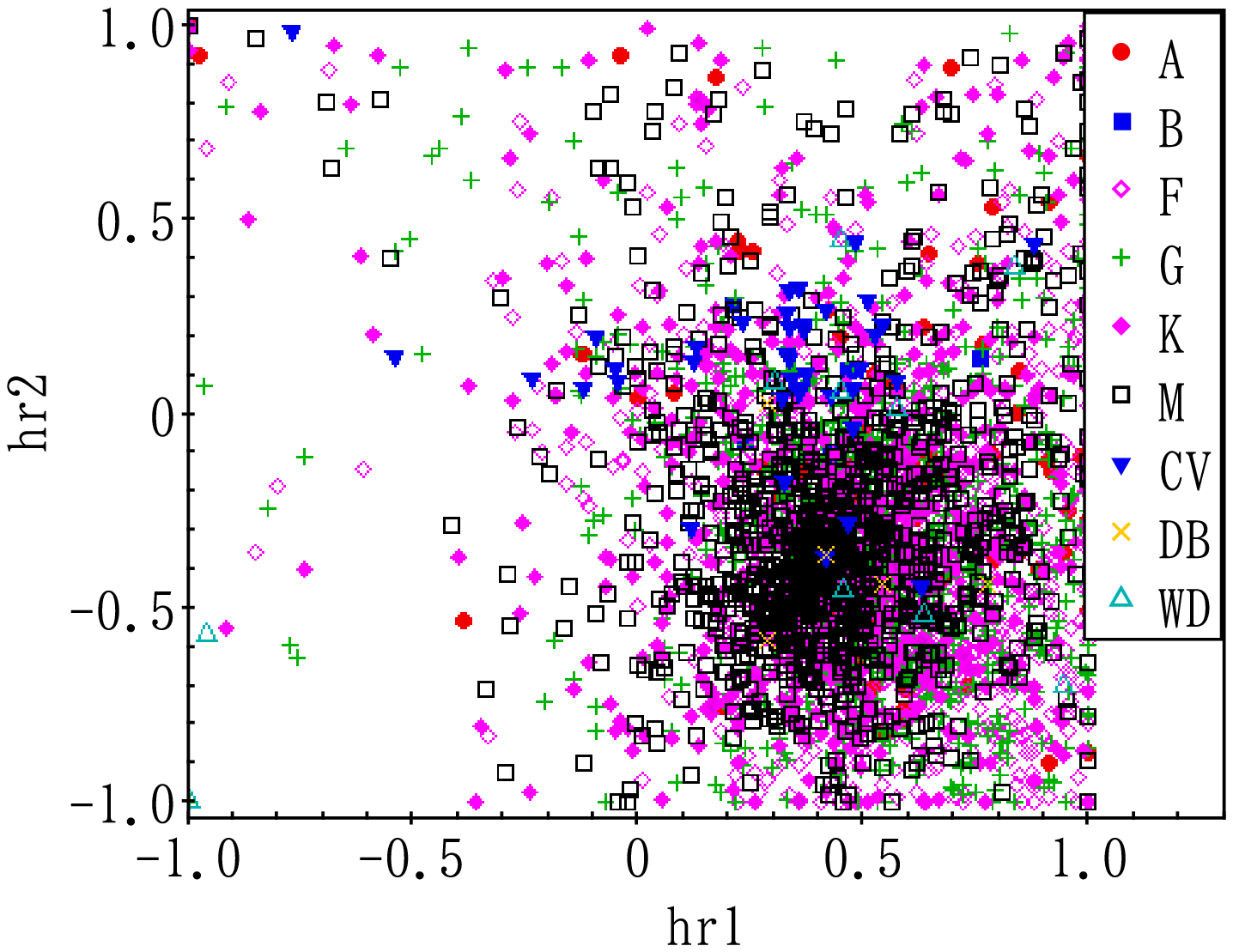}
\includegraphics[bb=-1 -1 427 356,width=5cm]{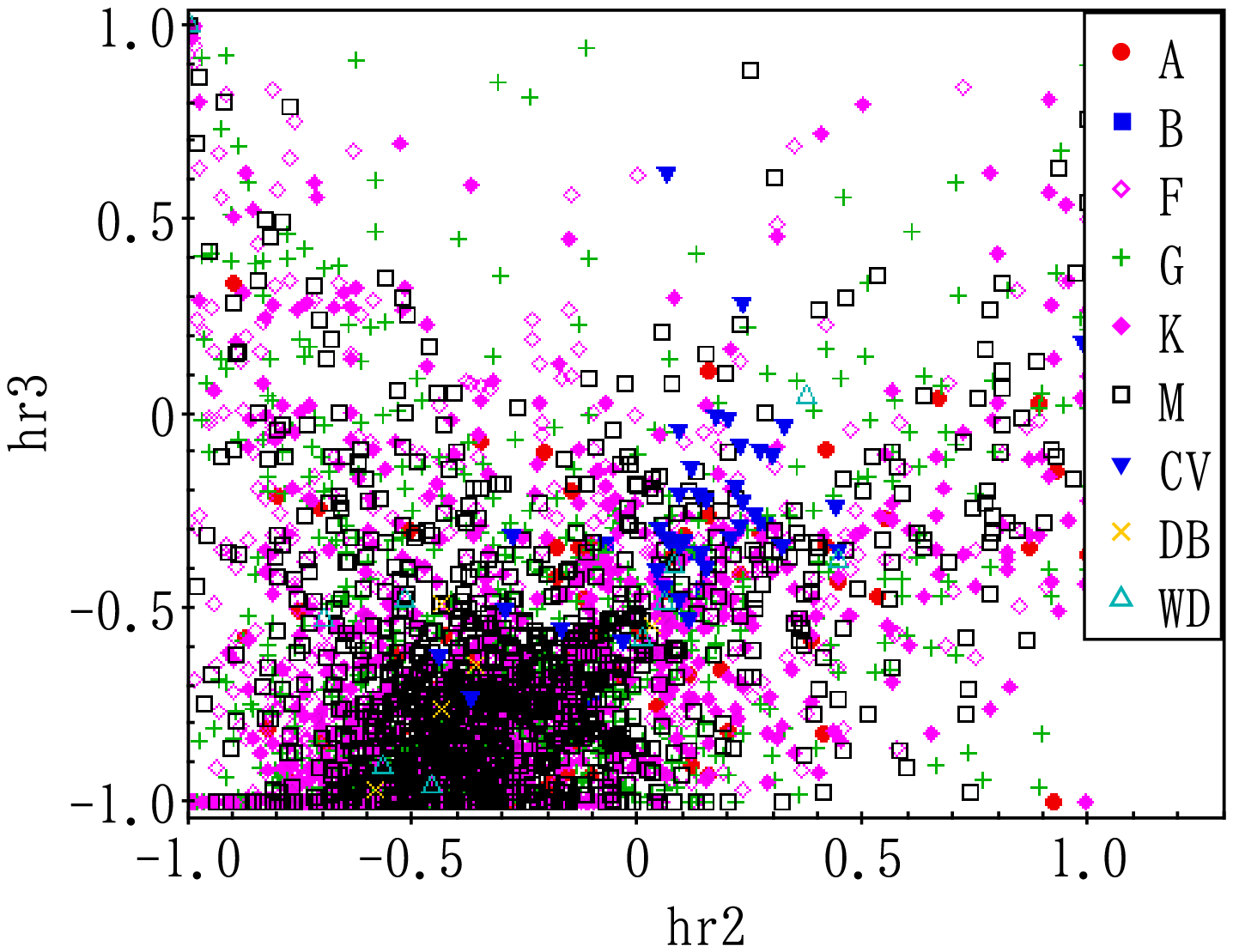}
\includegraphics[bb=-1 -1 427 356,width=5cm]{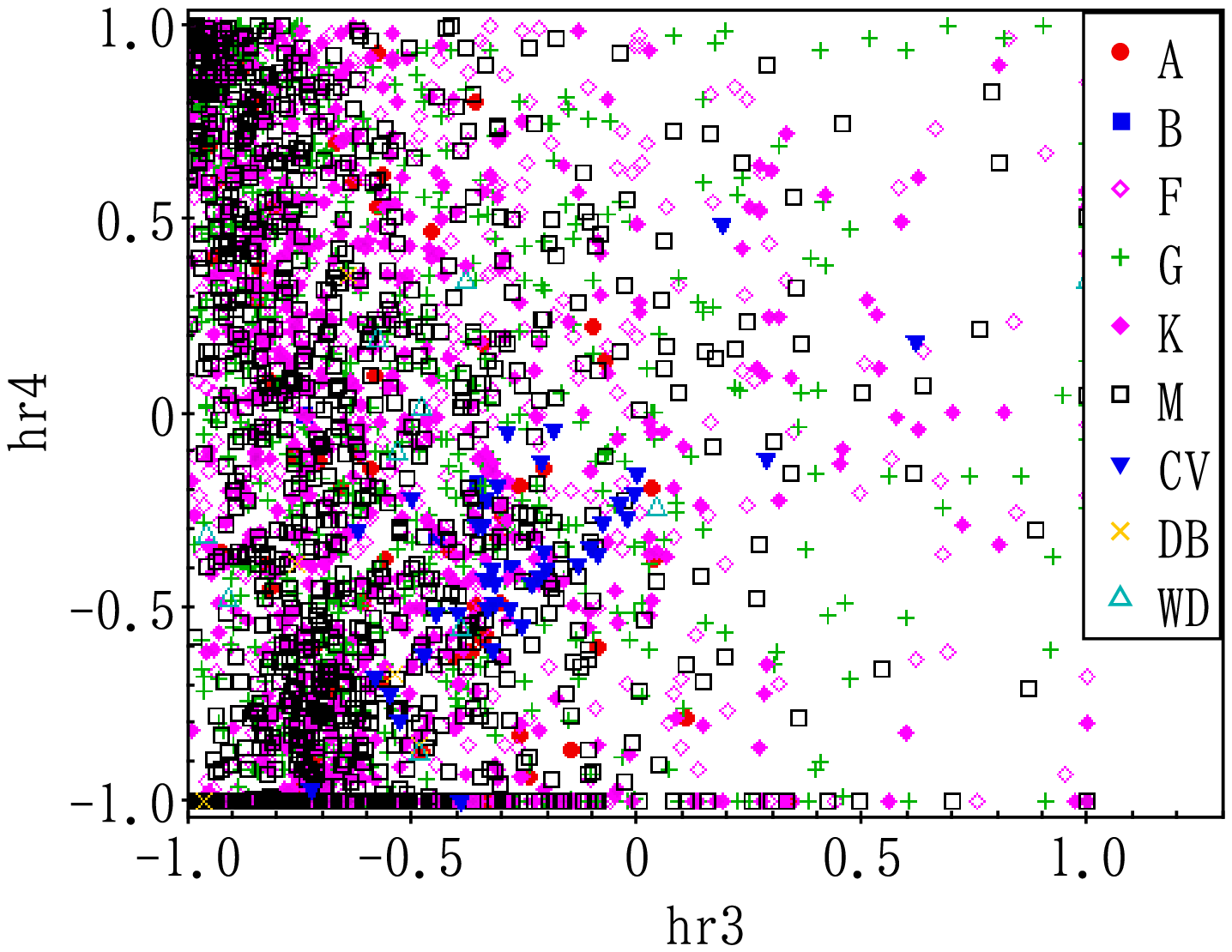}
\includegraphics[bb=-1 -1 427 356,width=5cm]{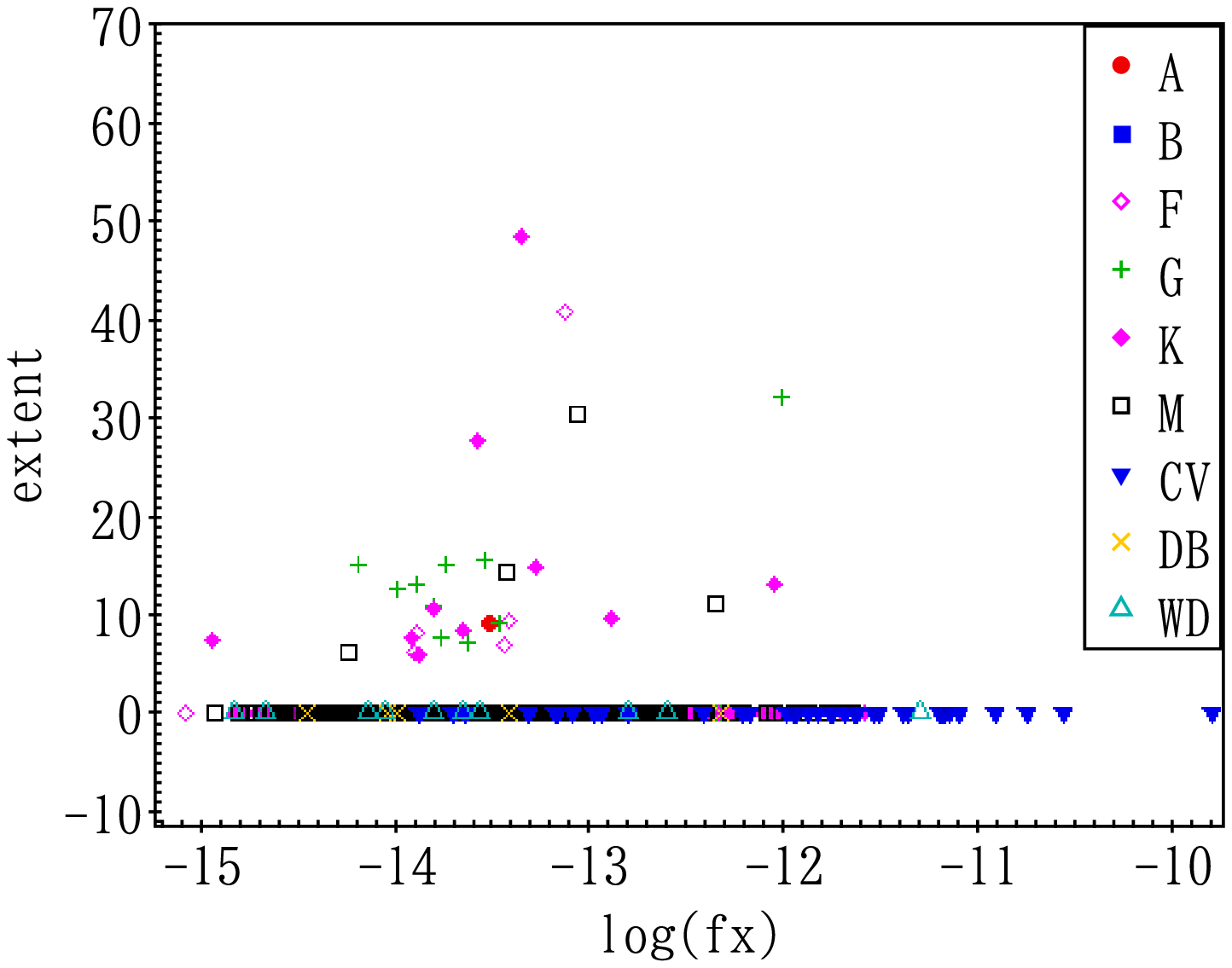}
\includegraphics[bb=-1 -1 427 356,width=5cm]{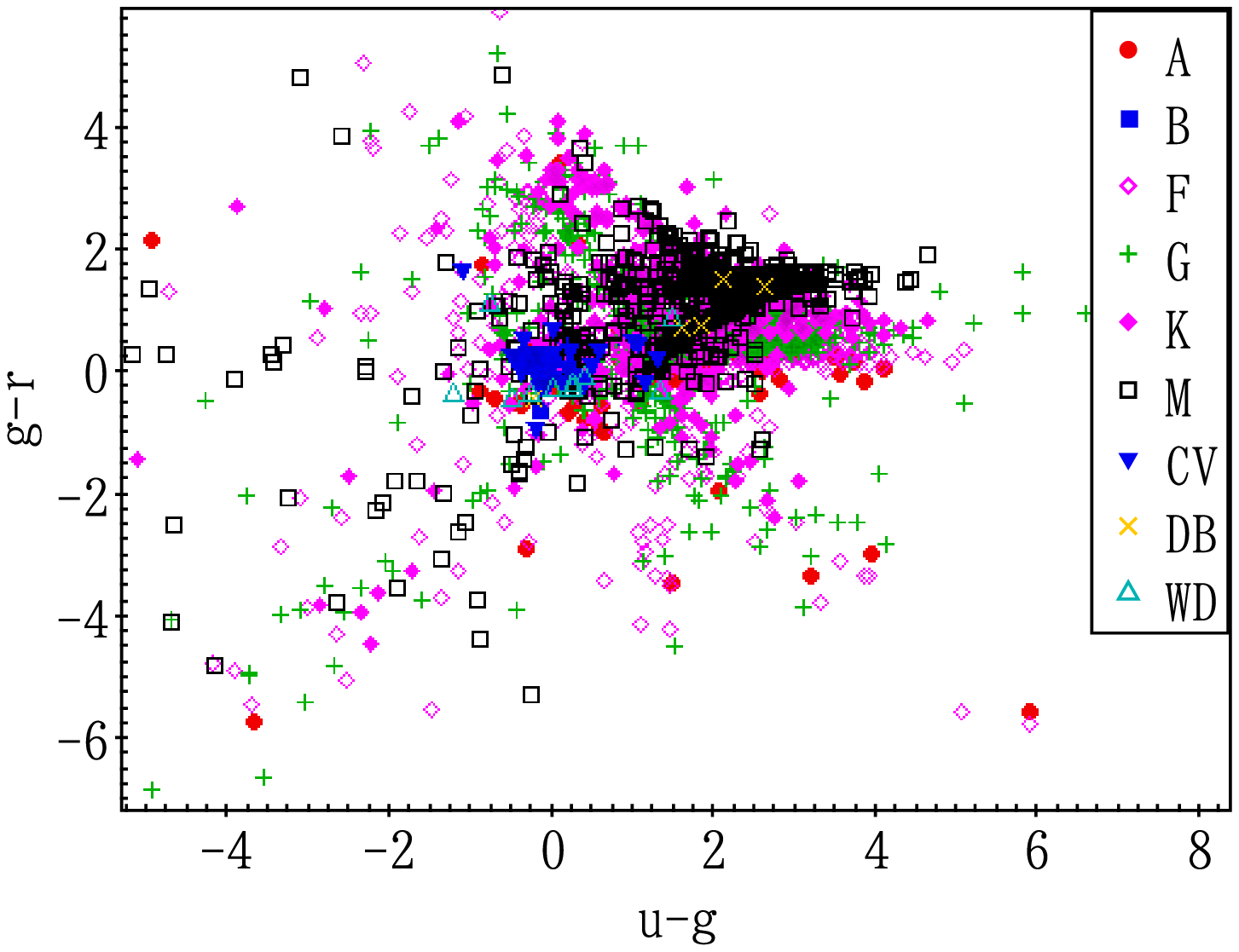}
\includegraphics[bb=-1 -1 427 356,width=5cm]{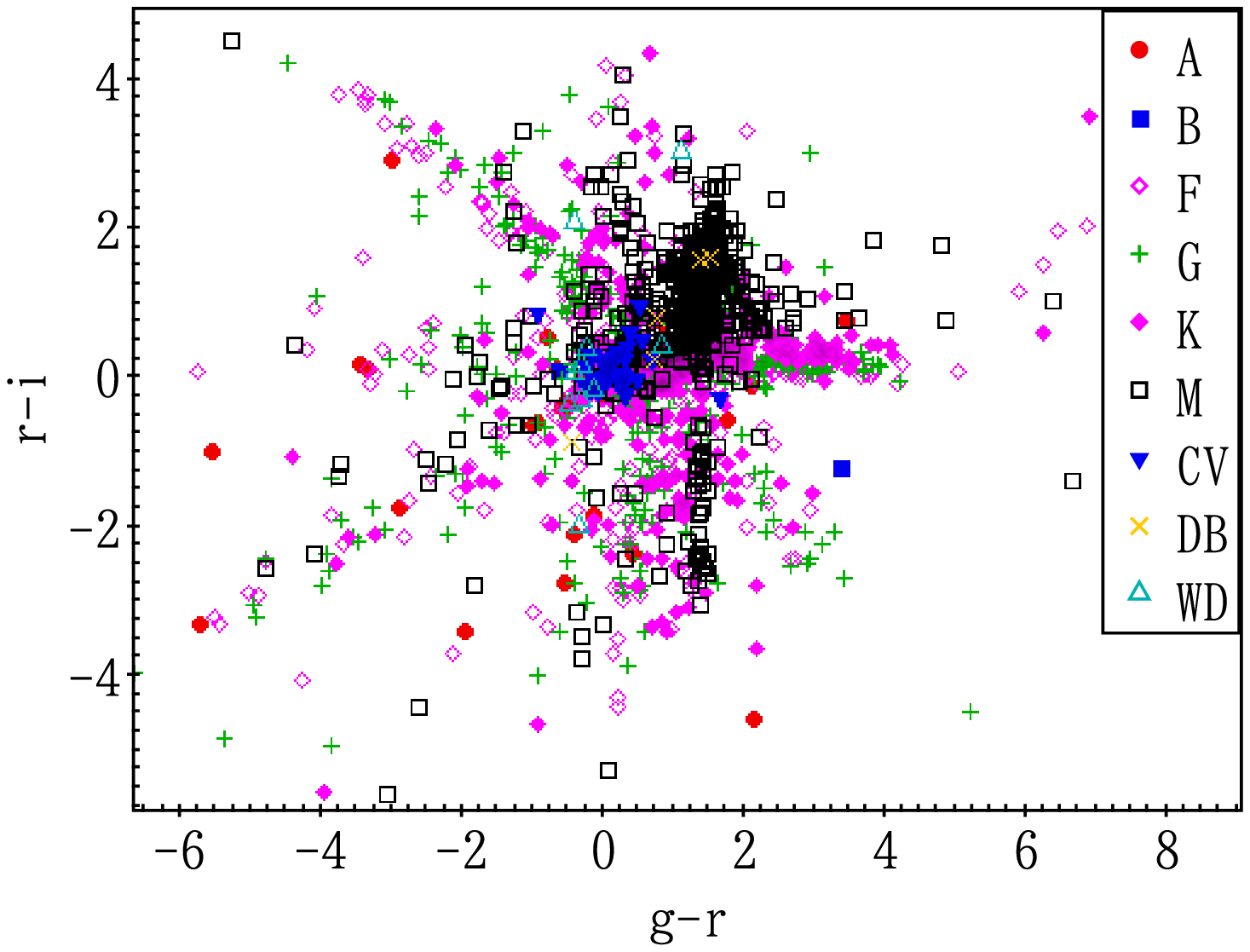}
\includegraphics[bb=-1 -1 427 356,width=5cm]{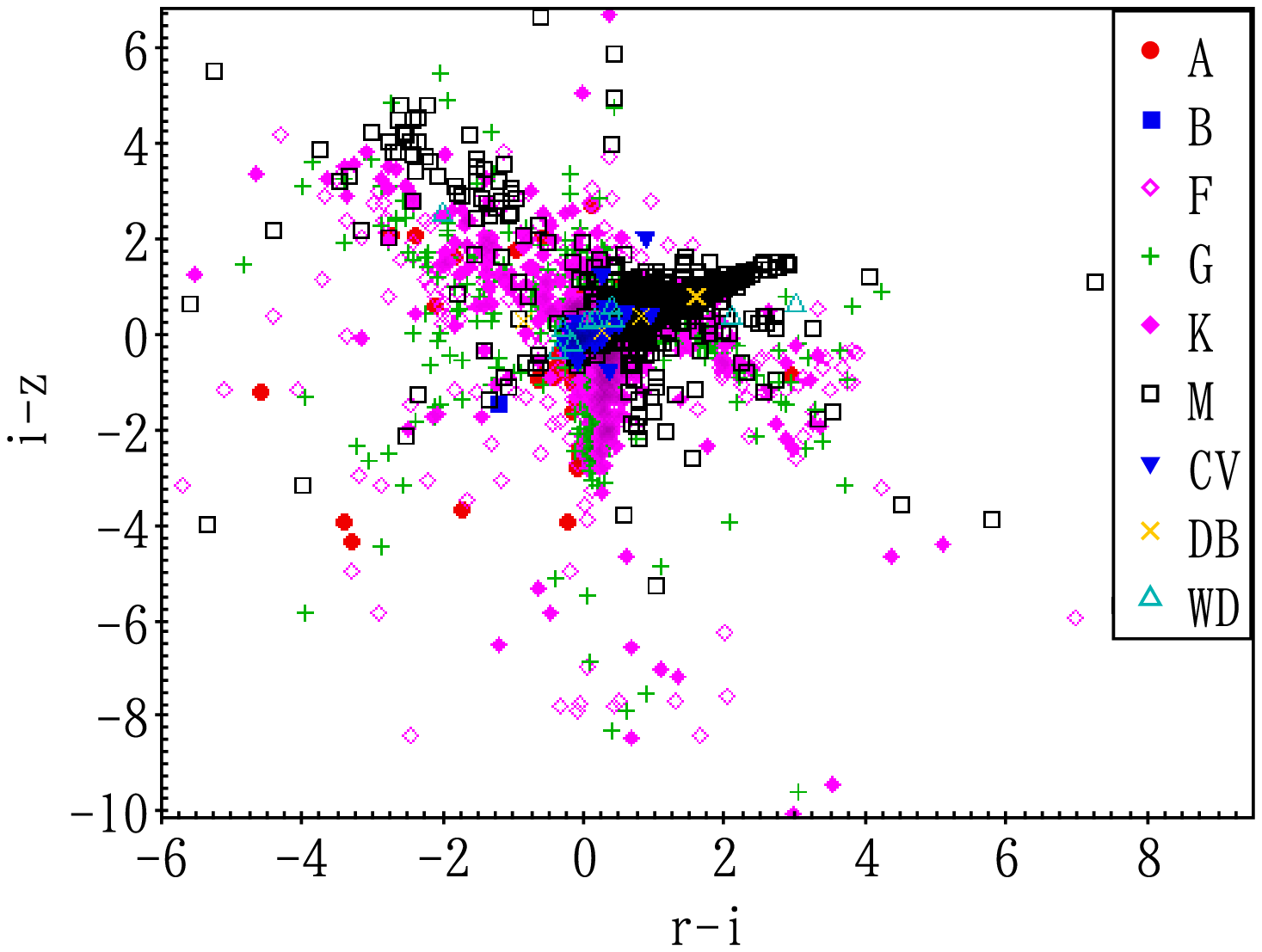}
\includegraphics[bb=-1 -1 427 356,width=5cm]{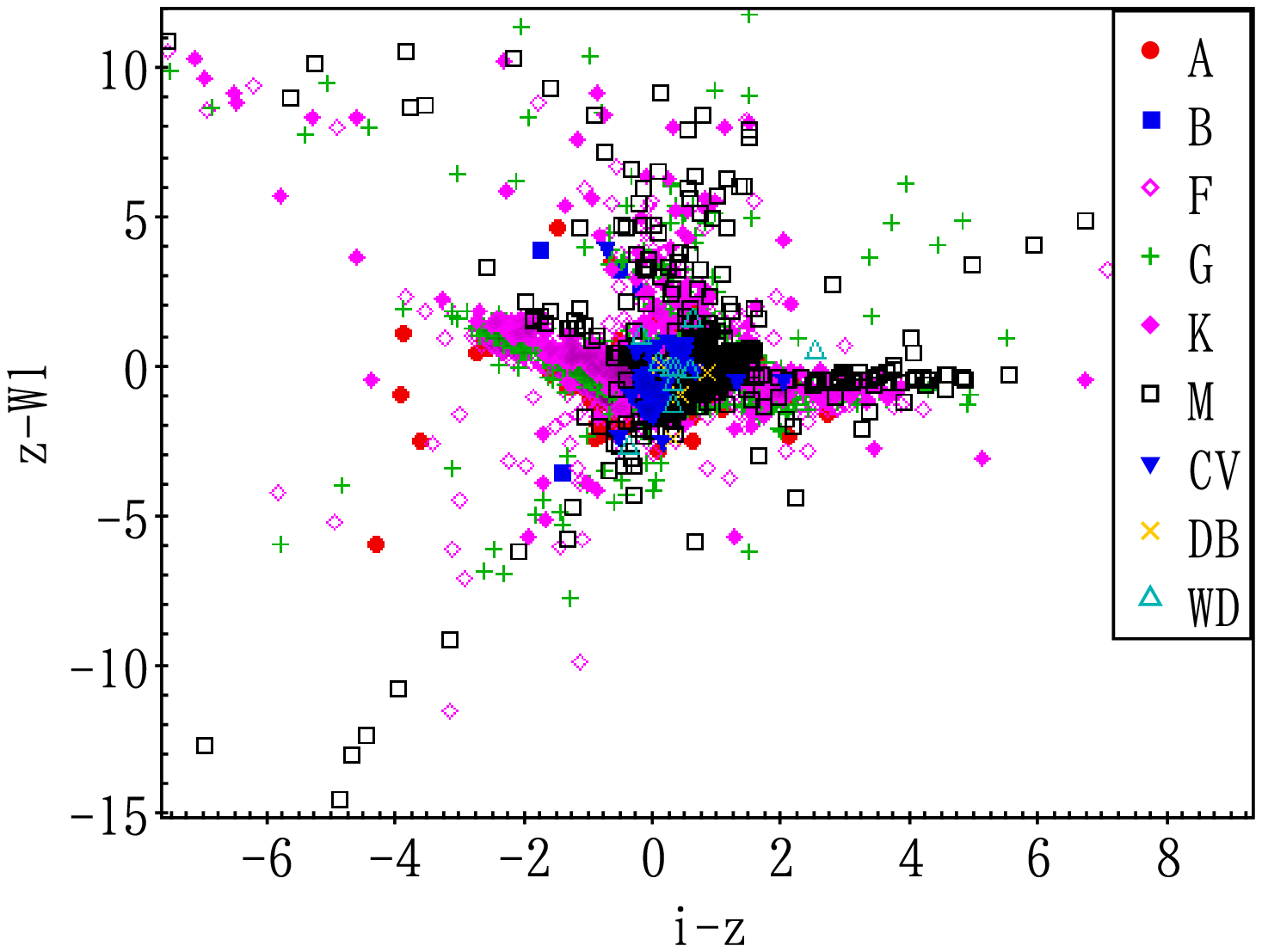}
\includegraphics[bb=-1 -1 427 356,width=5cm]{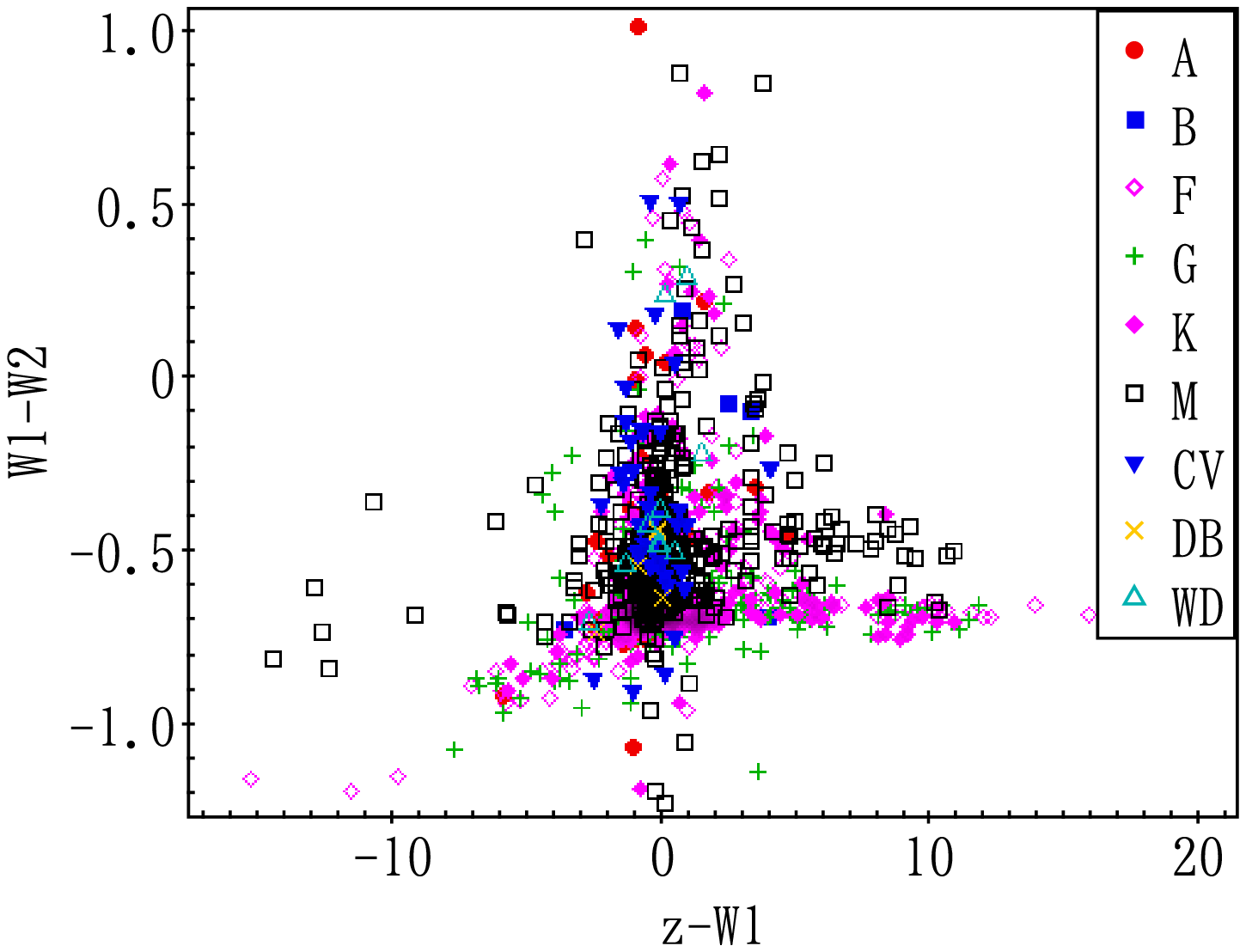}
\includegraphics[bb=-1 -1 427 356,width=5cm]{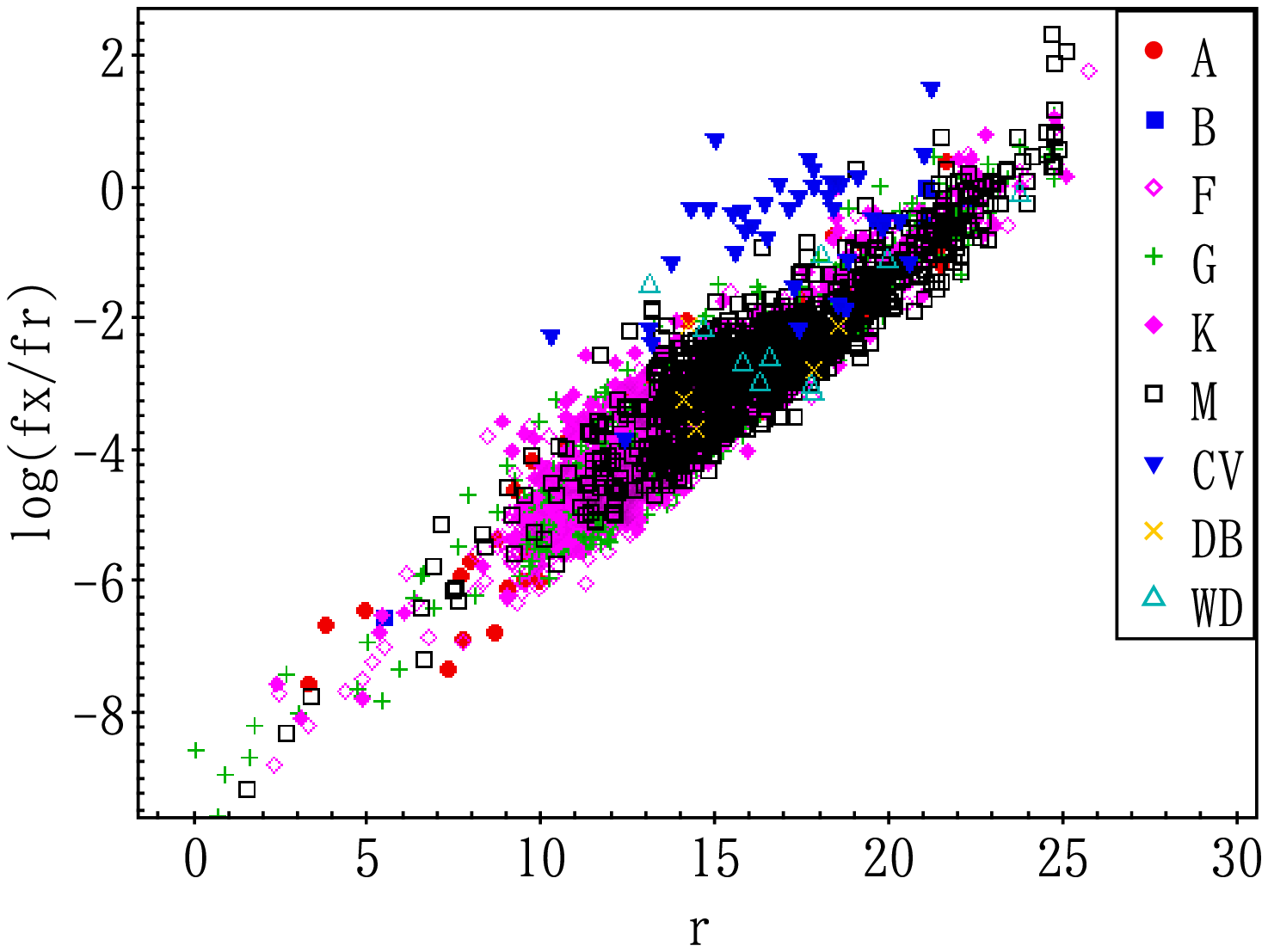}
\includegraphics[bb=-1 -1 427 356,width=5cm]{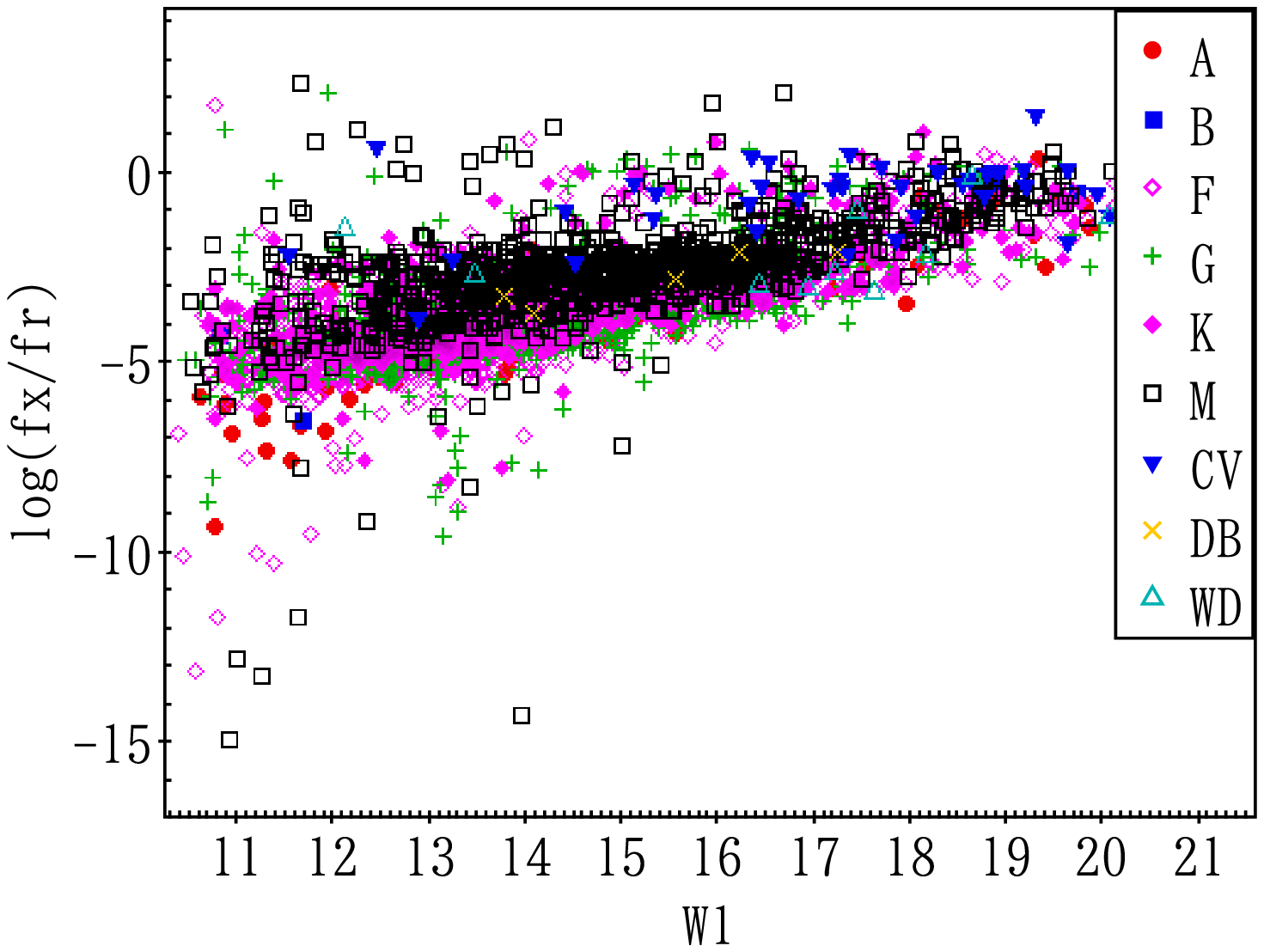}
\includegraphics[bb=-1 -1 427 356,width=5cm]{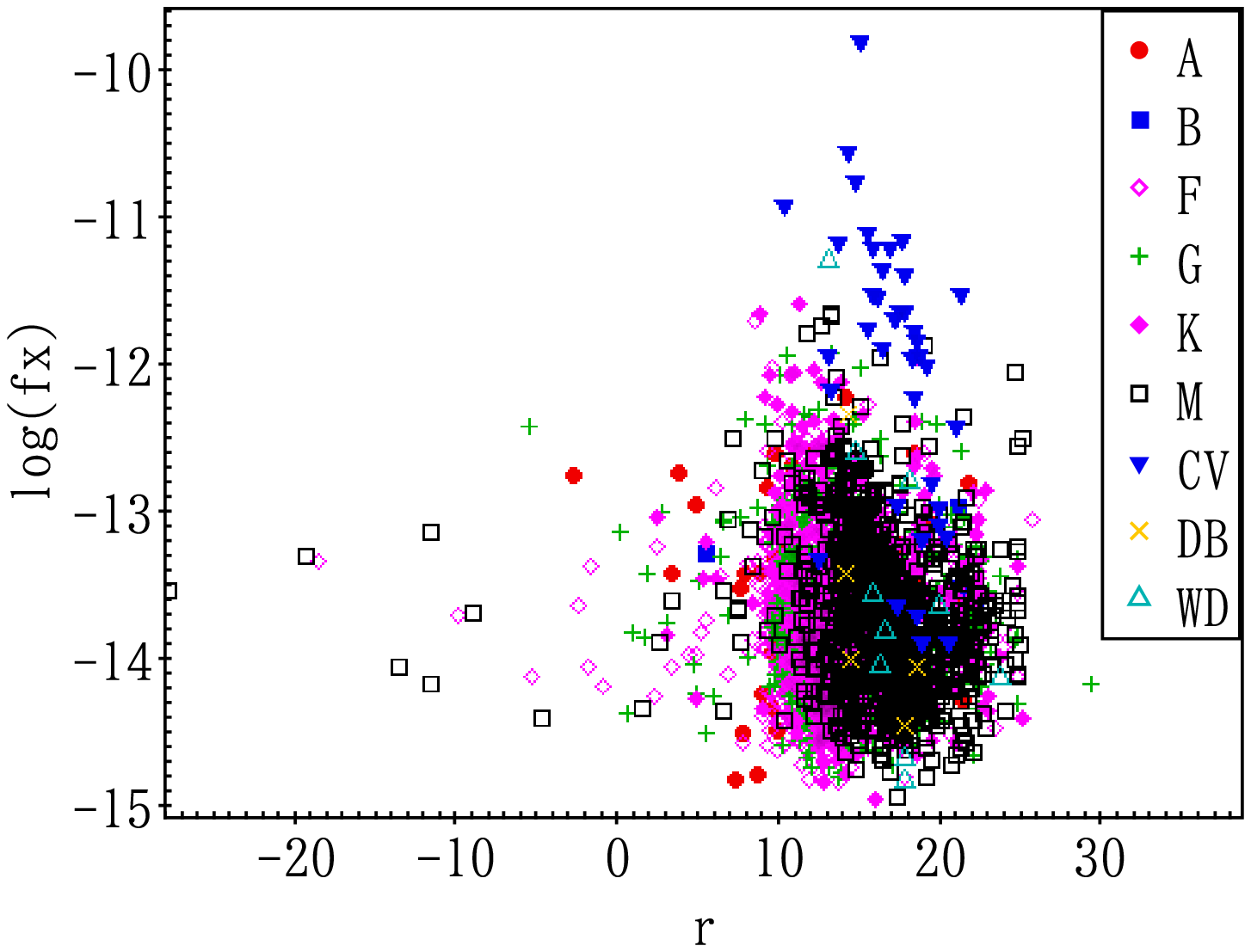}
\includegraphics[bb=-1 -1 427 356,width=5cm]{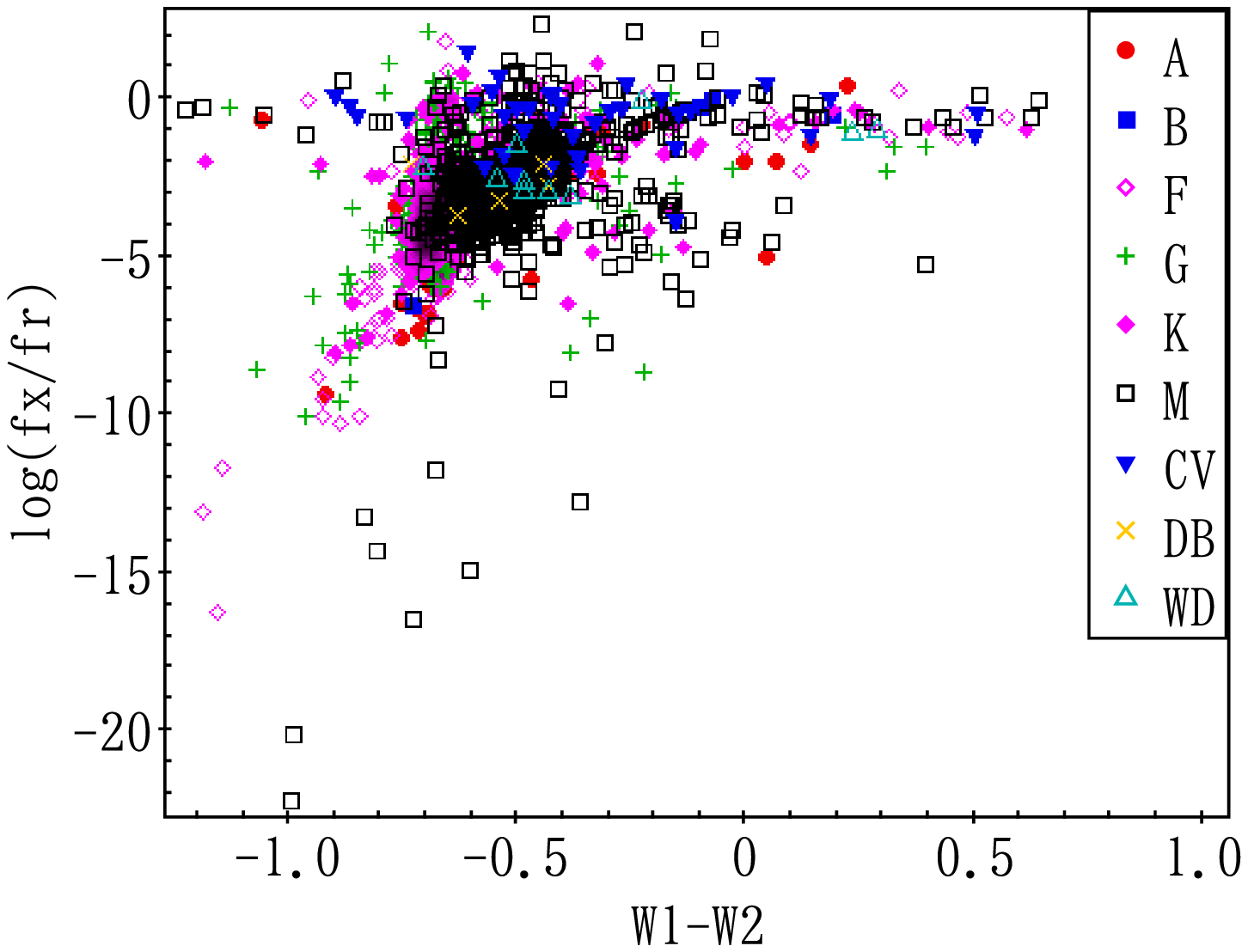}
\includegraphics[bb=-1 -1 427 356,width=5cm]{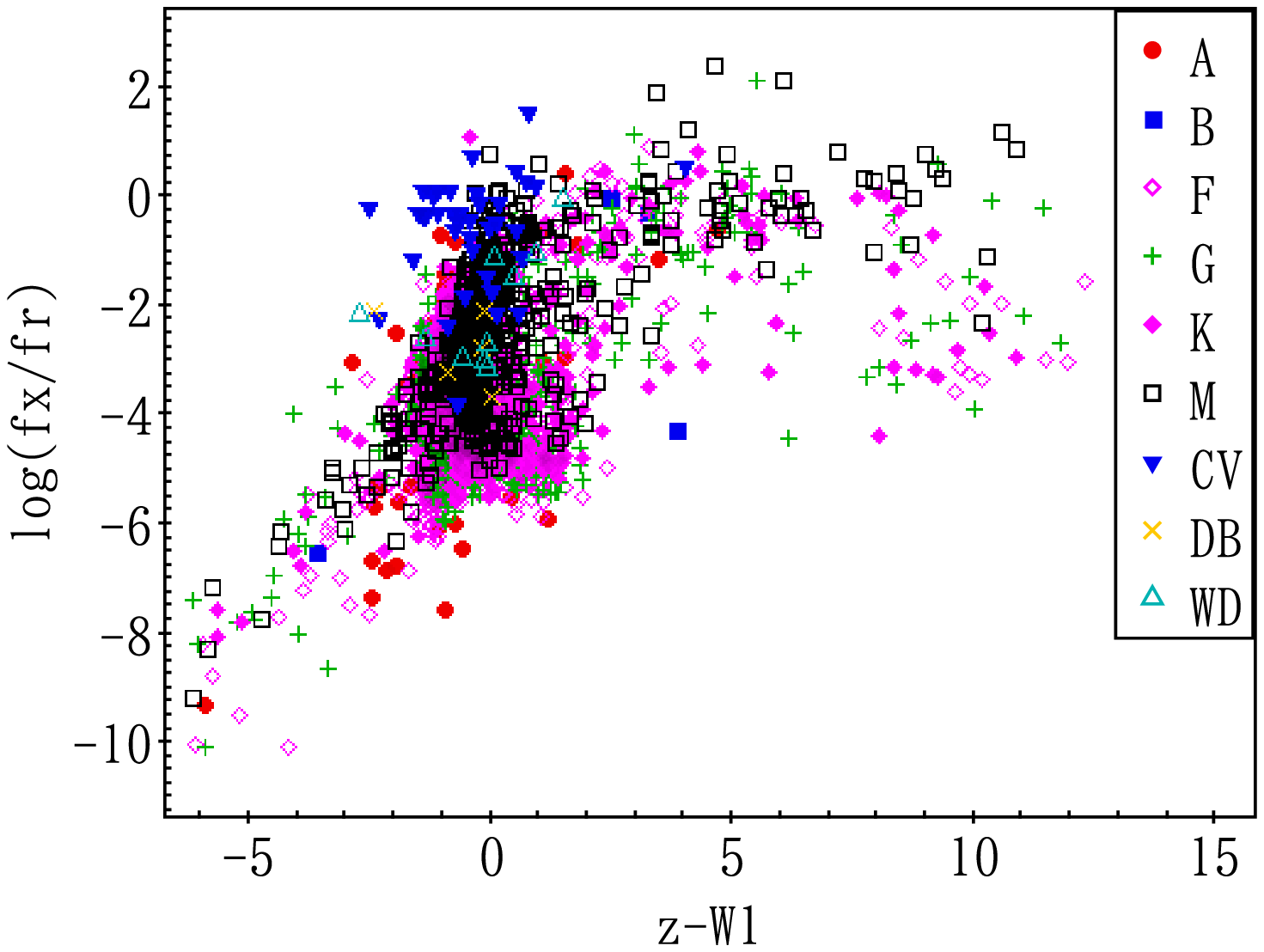}
\caption{The distribution of different spectral classes of stars in 2-d spaces, red filled circles for A stars, blue filled squares for B stars, purple open diamonds for F stars, green plus for G stars, purple filled diamonds for K stars, black open squares for M stars, blue filled down triangles for CV stars, yellow cross for double stars (DB), light green open up triangles for white dwarf stars (WD). }
\label{fig3}
\end{figure*}

\section{The method} \label{sec:method}

WEKA (The Waikato Environment for Knowledge Analysis; \citealt{Wit05}) is an open source software, which is effectively used for various machine learning tasks. It is implemented through a graphical user interface, standard terminal applications, or through a Java API. It is widely used for teaching, research, and industrial applications, and contains a plethora of built-in tools for standard machine learning tasks. These tasks include data pre-processing, classification, regression, clustering, association rules, attribute selection, and visualization realized by different algorithms. This software makes it easy to work with big data, perform and compare various machine learning algorithms. It is successfully applied in astronomy \citep{Zhao08,Zhang08,Zheng08}.

We try various classification algorithms provided by WEKA on our samples and only keep the better classification results. When performing the software, we all adopt the default setting by 10-fold validation while training a model. 10-fold validation refers to the dataset that is randomly divided into 10 parts, of which nine parts consist for training and one part remains for testing, this procedure is repeated 10 times.

The metrics commonly used to evaluate the performance of a classifier include Accuracy, Precision, Recall and F-measure. For a dataset, Accuracy is the ratio of the total number of correct predictions to the total number of predictions, Precision (also named as efficiency) is the fraction of true positive predictions among all true positive examples, Recall (also named as completeness) is the fraction of true positive predictions among all predicted positive examples, F-Measure is the weighted average of Precision and Recall.

\begin{equation}
 \mathrm{Accuracy} = \frac{TP+TN}{TP + TN + FP + FN}
\end{equation}
Here TP is the true positive sample, TN is the true negative sample, FP is the false positive sample, FN is the false negative sample.

\begin{equation}
 \mathrm{Precison} = \frac{TP}{TP + FP}, \mathrm{Recall} = \frac{TP}{TP + FN}
\end{equation}

\begin{equation}
 \mathrm{F-measure} = \frac{2\times Precision \times Recall}{Precision + Recall}
 \end{equation}

\subsection{Random forest}
Random forest is a supervised learning algorithm, which builds a randomized decision tree in each iteration of the bagging algorithm and gives impressive results with very large ensembles \citep{Bre01}. The bagging algorithm is applied to improve accuracy by reducing the variance to make the model more general and avoiding overfitting. For bagging, multiple subsets is taken as the training set. For each subset, a model created by the same algorithm is used to predict the output for the same test set. Averaging predictions is considered as the final prediction output. To further understand how the bagging algorithm works, we assume there are $N$ models and a Dataset. This dataset is split into training set and test set. Taking a sample of records from the training set, we train the first model with it. Then taking another sample from the training set, we train the second model with it. The similar process will be repeated for the $N$ number of models. Based on all predictions of $N$ models on the same test set, we adopt the model averaging technique like weighted average, variance or max voting to obtain the final prediction. Ensembles are a divide-and-conquer approach used to improve performance. For ensemble methods, ``weak learners" are grouped to form a ``strong learner". Each classifier, individually, is a ``weak learner" (base learner) while all the classifiers taken together are a ``strong learner". In a decision tree, the input data are separated into smaller and smaller sets from the tree root to its leaves. Random Forest creates many decision trees. When classifying a new object, each decision tree provides a classification. The final class of this object depends on the most votes among all the trees in the forest. This simplified random forest is shown in Figure~4. The advantage of using random forest is that it is able to deal with unbalanced and missing data and runs relatively fast.

\begin{figure}
\centering
\includegraphics[bb=0 0 536 467,width=8cm]{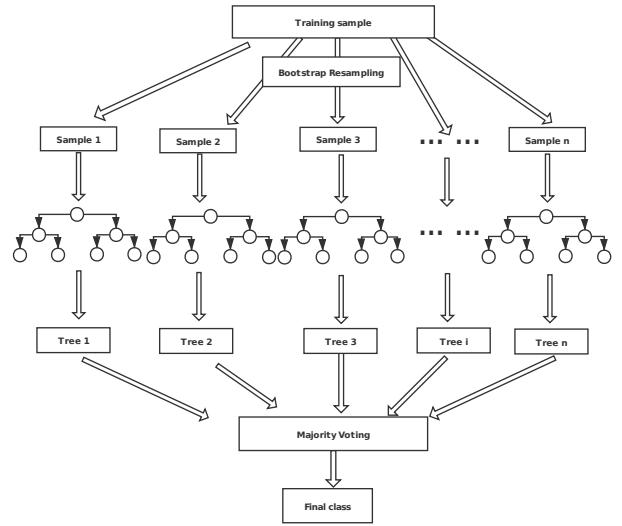}
\caption{The simplified random forest.}
\label{fig4}
\end{figure}

\subsection{Rotation forest}

Rotation forest is a powerful tree-based ensemble method based on feature extraction and designed to work with a smaller number of ensembles, and focuses on building accurate and diverse classifiers \citep{Rod06}. Feature extraction by Principal Component Analysis (PCA) is performed on $K$ subsets randomly split from the feature set in turn, here $K$ is a parameter of Rotation forest. All principle components are kept for each subset. Original data is handled by the principle component transformation and then used for training each base classifier. Its diversity is realized by the feature extraction carried out on each base classifier and its accuracy is ensured by all principal components kept and the whole data as training sample for each base classifier. Decision trees are usually selected because they are easily influenced by rotation of the feature axes. The difference between random forest and rotation forest is that rotation forest performs PCA on the feature subset to rebuild full feature space and achieves similar or better performance with fewer trees than by random forest. The detailed principle of rotation forest refers to \citet{Rod06}.

\subsection{LogitBoost}

LogitBoost is a boosting classification algorithm, based on the logistic regression method by minimizing the logistic loss \citep{Frid00}. Because noise and outliers exist in data and exponential loss function is used in LogitBoost, issues like overfitting will reduce a model accuracy. However classification errors are changed linearly instead of exponentially, thus this may improve the model accuracy and noise immunity. Here the Logitboost classification algorithm is trained using random forests as weak learners.

\section{Performance of the algorithms} \label{sec:performance}

We classify the X-ray sources into some subclasses of galaxies, stars and quasars, based on the input pattern of log$(f_x),hr1,hr2,hr3,hr4,extent,r,W1,u-g,g-r,r-i,i-z,z-W1,W1-W2,$log$(f_x/f_r)$). Since the LAMOST database hasn't given the subclassification of galaxies, we don't consider the galaxies from LAMOST when performing multi-classification. The subclasses of AGN and AGN BL are labelled as AGN, SB and SB BL labelled as SB, SF and SF BL labelled as SF, default value labelled as galaxies. LogitBoost is applied on the known sample without the galaxies from LAMOST by 10-fold validation. The classified result is described in Table~4. As shown in Table~4, the total accuracy adds up to 90.04 per~cent, the metrics of stars and quasars are above 92.7 per~cent while those of galaxies are unsatisfactory. The subclasses of galaxies are easily confused. The subclass of default value for galaxies assigned as galaxy belongs to normal galaxies, while the subclasses of AGN, SF, SB and BL belong to active galaxies. All metrics of normal galaxies are larger than 77.0 per~cent while those of active galaxies range from 7.8 per~cent to 76.6 per~cent. Active galaxies are inclined to be classified as normal galaxies or quasars. Obviously it is very difficult to discriminate active galaxies from the whole sample. Therefore we use the known samples from LAMOST and SDSS, and only classify the sample into galaxies, stars and quasars in the following work.

\begin{table*}
\begin{center}
\tiny
\caption[]{The performance of LogitBoost for multi-classification \label{tab:multiclass}}
 \begin{tabular}{rccccccccc|ccc}
 \hline
 \hline\noalign{\smallskip}
known$\downarrow$classified$\to$   &  AGN  &BL& SB&SF&galaxy&QSO&STAR&Precision&Recall&F-Measure\\
\hline
AGN     &149&8&7&98&269&177&10&50.0\%&20.8\%&29.3\%\\
BL      &12&22&1&9&193&34&10&44.9\%&7.8\%&13.3\% \\
SB      &4&0&141&68&23&141&18&76.6\%&35.7\%&48.7\%\\
SF      &53&3&28&472&251&205&42&58.9\%&44.8\%&50.9\%\\
galaxy  &45&14&4&100&3 698&605&70&77.0\%&81.5\%&79.2\%\\
QSO     &35&0&0&50&268&20 657&30&94.0\%&98.2\%&96.1\%\\
STAR    &0&2&3&4&103&148&3 298&94.8\%&92.7\%&93.7\%  \\
\hline
Total Accuracy&&&&&90.04\%&&&&&\\
\hline
\end{tabular}
\end{center}
\end{table*}

For the sample only from X-ray band, the classification performance of random forest and rotation forest is shown in Table~5. The input pattern for this sample is log$(f_x), hr1, hr2, hr3, hr4, extent$. As shown in Table~5, only for galaxies, Recall and F-Measure decrease, but for stars and quasars, all metrics increase, comparing the performance of rotation forest with random forest. Rotation forest outperforms random forest in terms of Accuracy (77.80 per~cent vs. 77.46 per~cent). Only with information from X-ray band, the classification metrics of quasars are satisfying while those of galaxies and stars are not good when considering Precision, Recall and F-measure.

\begin{table*}
\begin{center}
\tiny
\caption[]{The performance of random forest and rotation forest with log$(f_x), hr1, hr2, hr3, hr4, extent$. \label{tab:comp2}}
 \begin{tabular}{r|ccc|ccc}
 \hline
 \hline\noalign{\smallskip}
 Method   &             &random forest&            &              &rotation forest&\\
\hline
Class     &Precision &Recall  &F-Measure&Precision &Recall  &F-Measure\\
 \hline\noalign{\smallskip}
QSO   &82.1\%  &93.4\%  &87.4\%& 81.4\%  &94.9\%  &87.6\% \\
GALAXY&63.0\%  &43.4\%  &51.4\%& 66.0\%  &40.4\%  &50.1\%  \\
STAR  &64.0\%  &52.4\%  &57.6\%& 65.1\%  &52.7\%  &58.3\%  \\
\hline
Total Accuracy&&77.46\% &    &       &77.80\%&\\
\hline
\end{tabular}
\end{center}
\end{table*}

For the sample from X-ray and optical bands, the classification performance of random forest and LogitBoost is indicated in Table~6. The input pattern is log$(f_x), hr1, hr2, hr3, hr4, extent, r, u-g, g-r, r-i, i-z,$ log$(f_x/f_r)$. As indicated in Table~6, all metrics for LogitBoost are better than those for random forest, and all of them are higher than 84.8 per~cent. Only touching on quasars and stars, the metrics are above 87.5 per~cent. LogitBoost is superior to random forest for this case, as its accuracy amounts to 92.82 per~cent.

\begin{table*}
\begin{center}
\tiny
\caption[]{The performance of random forest and LogitBoost with log$(f_x), hr1, hr2, hr3, hr4, extent,r,u-g,g-r,r-i,i-z,$log$(f_x/f_r)$. }
 \begin{tabular}{r|ccc|ccc}
 \hline
 \hline\noalign{\smallskip}
  Method   &             &random forest&            &              &LogitBoost&\\
\hline
Class     &Precision &Recall  &F-Measure&Precision &Recall  &F-Measure\\
 \hline\noalign{\smallskip}
QSO   &94.4\%  &96.1\%  &95.2\%  &94.5\%  &96.2\%  &95.4\%\\
GALAXY&85.6\%  &84.8\%  &85.2\%  &86.1\%  &85.3\%  &85.7\%\\
STAR  &95.9\%  &87.5\%  &91.5\%  &96.2\%  &88.1\%  &92.0\%\\
\hline
Total Accuracy&&92.57\%&&&92.82\%&\\
\hline
\end{tabular}
\end{center}
\end{table*}

For the sample from X-ray and infrared bands, the classification performance of random forest and LogitBoost is described in Table~7. The input pattern is
log$(f_x), hr1, hr2, hr3, hr4, extent, W1, W1-W2$. As depicted in Table~7, the performance of random forest is a little better than LogitBoost in terms of total accuracy. All metrics for random forest are near to those of LogitBoost. The accuracy of galaxies is still worse than that of quasars and stars. Nevertheless, all metrics are better than 76.1 per~cent. The total accuracy of random forest is 89.42 per~cent.

\begin{table*}
\begin{center}
\tiny
\caption[]{The performance of random forest and LogitBoost with log$(f_x), hr1, hr2, hr3, hr4, extent, W1, W1-W2$. }
 \begin{tabular}{r|ccc|ccc}
 \hline
 \hline\noalign{\smallskip}
  Method   &             &random forest&            &              &LogitBoost&\\
\hline
Class     &Precision &Recall  &F-Measure&Precision &Recall  &F-Measure\\
 \hline\noalign{\smallskip}
QSO   &93.4\%  &95.9\%  &94.6\%  &93.4\%  &96.8\%  &95.9\%\\
GALAXY&79.1\%  &76.1\%  &77.5\%  &78.9\%  &76.1\%  &76.1\%\\
STAR  &85.1\%  &78.2\%  &81.5\%  &85.1\%  &82.3\%  &77.9\%\\
\hline
Total Accuracy&&89.42\%&&&89.38\%&\\
\hline
\end{tabular}
\end{center}
\end{table*}

For the sample from X-ray, optical and infrared bands, the classification performance of random forest and LogitBoost is listed in Table~8. The input pattern is log$(f_x),hr1,hr2,hr3,hr4,extent,r,W1,u-g,g-r,r-i,i-z,z-W1,W1-W2,$log$(f_x/f_r)$. As narrated in Table~8, even for galaxies, the metrics are greater than 87.1 per~cent; for stars, the metrics are above 90.7 per~cent; for quasars, the metrics are higher than 95.4 per~cent. All metrics except precision for LogitBoost are greater than those of random forest. Compared to random forest, LogitBoost has a slight advantage and its total accuracy adds up to 94.26 per~cent.

\begin{table*}
\begin{center}
\tiny
\caption[]{The performance of random forest and LogitBoost with log$(f_x),hr1,hr2,hr3,hr4,extent,r,W1,u-g,g-r,r-i,i-z,z-W1,W1-W2,$log$(f_x/f_r)$. }
 \begin{tabular}{r|ccc|ccc}
 \hline
 \hline\noalign{\smallskip}
  Method   &             &random forest&            &              &LogitBoost&\\
\hline
Class     &Precision &Recall  &F-Measure&Precision &Recall  &F-Measure\\
 \hline\noalign{\smallskip}
QSO   &95.4\%  &97.0\%  &96.2\%  &95.6\%  &97.1\%  &96.3\%\\
GALAXY&88.4\%  &87.1\%  &87.7\%  &89.1\%  &87.5\%  &88.3\%\\
STAR  &96.9\%  &90.7\%  &93.7\%  &96.8\%  &91.2\%  &93.9\%\\
\hline
Total Accuracy&&94.03\%&&&94.26\%&\\
\hline
\end{tabular}
\end{center}
\end{table*}

In order to check how the observational errors influence the performance of a classifier, we take the XMM-SDSS sample for example. Setting $\sigma_u<0.3$, $\sigma_g<0.3$, $\sigma_r<0.3$, $\sigma_i<0.3$ and $\sigma_z<0.3$, the known sample size changes from 31 800 to 26 428, the performance of random forest and LogitBoost is shown in Table~9. Comparing the result in Table~9 with that in Table~6, the performance of random forest and LogitBoost both improves with higher quality data (94.73 per~cent vs. 92.57 percent for random forest, 94.93 per~cent vs. 92.82 per cent for LogitBoost) in terms of accuracy. Although higher quality data leads to higher performance of a classifier, the number of sources with X-ray emission is small in nature. So we do not set magnitude error limitation on the samples in our work.

\begin{table*}
\begin{center}
\tiny
\caption[]{The performance of random forest and LogitBoost with log$(f_x), hr1, hr2, hr3, hr4, extent,r,u-g,g-r,r-i,i-z,$log$(f_x/f_r)$ when $\sigma_u<0.3$, $\sigma_g<0.3$, $\sigma_r<0.3$, $\sigma_i<0.3$ and $\sigma_z<0.3$. }

 \begin{tabular}{r|ccc|ccc}
 \hline
 \hline\noalign{\smallskip}
  Method   &             &random forest&            &              &LogitBoost&\\
\hline
Class     &Precision &Recall  &F-Measure&Precision &Recall  &F-Measure\\
 \hline\noalign{\smallskip}
QSO   &95.7\%  &98.1\%  &96.9\%  &95.9\%  &98.1\%  &97.0\%\\
GALAXY&88.8\%  &83.7\%  &86.2\%  &89.3\%  &84.1\%  &86.6\%\\
STAR  &96.4\%  &89.8\%  &93.0\%  &96.5\%  &90.9\%  &93.6\%\\
\hline
Total Accuracy&&94.73\%&&&94.93\%&\\
\hline
\end{tabular}
\end{center}
\end{table*}

\section{Discussion and application}
Comparing Tables~5-8, the worst result belongs to the sample only from X-ray band as expected. Adding the information from optical band and/or infrared band, the classification accuracy increases for any classifier, nevertheless the accuracy with X-ray and optical bands is better than that with X-ray and infrared bands. The best performance is obtained with all information from X-ray, optical and infrared bands. There is not any algorithm which shows the best performance for any dataset. For the sample from X-ray band, the rotation forest classifier is the best; for the sample from X-ray and infrared bands, random forest is superior to all other algorithms; while for another two samples, LogitBoost shows its superiority.

In reality, some X-ray sources have information from X-ray, optical and infrared bands, some have information from X-ray and infrared bands, some have information from X-ray and optical bands, even some only have X-ray information. Based on the known samples with spectral classes, we need to construct four classifiers for the four situations to predict the unknown X-ray sources. For the sources only with X-ray information, a rotation forest classifier is built with the known samples with spectral classes to predict their classes and probability. For the sources with X-ray and infrared bands, a random forest classifier is created with the known samples with spectral classes to predict their classes and probability. For the sources from X-ray and optical bands or from X-ray, optical and infrared bands, LogitBoost classifiers are constructed with the corresponding known samples with spectral classes to predict their classes and probability, respectively. For the 4XMM-DR9 sources, all predicted results are shown in Table~10. Table~10 provides the classification information for the 4XMM-DR9 Sources. The gained information is of great value for the further research on the characteristics and physics of X-ray sources.

\begin{table*}
\begin{center}
\tiny
\caption[]{Classification of 4XMM-DR9 sources.}
 \begin{tabular}{rcccccccccc}
 \hline
 \hline\noalign{\smallskip}
 srcid          &sc\_ra         & sc\_dec      &Class\_x  &P$_x$   &Class\_xo&P$_{xo}$&Class\_xi&P$_{xi}$&Class\_xio&P$_{xio}$\\
 \hline\noalign{\smallskip}
 200001101010001&64.9255899382624&55.9993455276706&GALAXY&0.718&STAR&1.0&&&&\\
 200001101010002&64.9714038006107&55.8049026564271&GALAXY&0.427&QSO&0.996&GALAXY&0.894&QSO&0.998\\
 200001101010003&65.0767247976311&55.9307646652894&GALAXY&0.456&QSO&0.963&GALAXY&1.0&STAR&0.722\\
 200001101010004&65.1112285547752&55.9955363739078&GALAXY&0.746&GALAXY&1.0&GALAXY&0.993&GALAXY&1.0\\
 200001101010005&64.996228987918&56.2248168838265&STAR&0.653&STAR&1.0&STAR&1.0&STAR&1.0\\
 200001101010006&64.9348515102436&55.9291776566485&GALAXY&0.506&&&&&&\\
 200001101010007&64.8232313435949&55.9849189955416&GALAXY&0.485&QSO&1.0&GALAXY&1.0&QSO&0.999\\
 200001101010008&65.0734121719342&55.9823011754657&QSO&0.491&STAR&1.0&STAR&0.939&STAR&1.0\\
 200001101010009&65.0167233356568&55.9421102139164&QSO&0.537&QSO&0.997&&&&\\
 200001101010010&64.9101008917805&56.0710218248335&QSO&0.502&&&&&&\\
 200001101010011&64.9050705152553&56.0644078750126&QSO&0.539&GALAXY&0.955&GALAXY&0.999&GALAXY&0.943\\
 200001101010012&65.2336693528087&55.8993466831422&GALAXY&0.479&&&STAR&0.986&&\\
 200001101010013&64.8914247247132&55.9585111145714&QSO&0.523&&&&&&\\
 200001101010014&64.6507013015166&56.0418886508129&QSO&0.517&STAR&0.999&GALAXY&0.986&STAR&1.0\\
 200001101010015&64.7925428495702&55.896051166999&STAR&0.572&STAR&1.0&STAR&1.0&STAR&1.0\\
 200001101010016&65.1527613793266&55.9300031359814&QSO&0.489&&&GALAXY&0.999&&\\
 200001101010017&65.0424438892887&56.1513807784794&QSO&0.488&QSO&1.0&&&&\\
 200001101010018&64.725085117142&55.891398599223&GALAXY&0.579&STAR&1.0&GALAXY&1.0&STAR&0.996\\
 200001101010019&65.1553866221519&55.8977634034868&GALAXY&0.452&GALAXY&1.0&STAR&0.654&STAR&0.702\\
 200001101010020&64.9468670845107&55.9626521430764&GALAXY&0.657&GALAXY&0.981&GALAXY&0.998&GALAXY&0.98\\

\hline
\end{tabular}
\\
(Class\_x means classification and P$_x$ is their classification probabilities from X-ray band; Class\_xo means classification and P$_{xo}$ is their classification probabilities from X-ray and optical bands; Class\_xi means classification and P$_{xi}$ is their classification probabilities from X-ray and infrared bands; Class\_xio means classification and P$_{xio}$ is their classification probabilities from X-ray, infrared and optical bands.\\ This whole table is available in the website http://paperdata.china-vo.org/zyx/table10.csv. A portion is shown here for guidance about its form and content.)
\end{center}
\end{table*}

% This table is available in its entirety in a machine-readable form in the online journal. A portion is shown here for guidance regarding its form and content.

\section{Conclusions} \label{sec:conclusions}
Based on the distribution of stars, galaxies and quasars in 2-d spaces, it is difficult to discriminate them and their subclasses clearly. Similarly, given the distribution of all spectral classes of stars in 2-d spaces, it is also not so easy to separate them. But CV stars and M stars are easily to stand out in some 2-d spaces. Of the entire X-ray sample, quasars occupy the majority while stars and galaxies only cover a minority. With X-ray information and spectral classes of known X-ray sources, we create a rotation forest classifier to assign classification results and their probabilities for all 4XMM-DR9 sources. Based on information from X-ray and infrared bands as well as spectral classes of known X-ray sources, a random forest classifier is used to discriminate X-ray sources. By means of properties from X-ray, optical and/or infrared bands and spectral classes of known X-ray sources, we build LogitBoost classifiers to predict X-ray sources. The predicted results from different methods with different input properties are listed in an entire table, which may be used to further study X-ray properties of various kinds of objects in detail.

\section{Acknowledgements}
We are very grateful to the referees for their constructive suggestions. This paper is funded by the National Natural Science Foundation of China under grants No.11873066 and No.U1731109. This research has made use of data obtained from the 4XMM \emph{XMM-Newton} serendipitous source catalogue compiled by the 10 institutes of the \emph{XMM-Newton} Survey Science Centre selected by ESA. This publication makes use of data products from the Wide-field Infrared Survey Explorer, which is a joint project of the University of California, Los Angeles, and the Jet Propulsion Laboratory/California Institute of Technology, funded by the National Aeronautics and Space Administration. The Guoshoujing Telescope (the Large Sky Area Multi-object Fiber Spectroscopic Telescope, LAMOST) is a National Major Scientific Project built by the Chinese Academy of Sciences. Funding for the project has been provided by the National Development and Reform Commission. LAMOST is operated and managed by the National Astronomical Observatories, Chinese Academy of Sciences.

We acknowledgment SDSS databases. Funding for the Sloan Digital Sky Survey IV has been provided by the Alfred P. Sloan Foundation, the U.S. Department of Energy Office of Science, and the Participating Institutions. SDSS-IV acknowledges support and resources from the Center for High-Performance Computing at the University of Utah. The SDSS web site is www.sdss.org. SDSS-IV is managed by the Astrophysical Research Consortium for the Participating Institutions of the SDSS Collaboration including the Brazilian Participation Group, the Carnegie Institution for Science, Carnegie Mellon University, the Chilean Participation Group, the French Participation Group, Harvard-Smithsonian Center for Astrophysics, Instituto de Astrof\'isica de Canarias, The Johns Hopkins University, Kavli Institute for the Physics and Mathematics of the Universe (IPMU) /University of Tokyo, Lawrence Berkeley National Laboratory, Leibniz Institut f\"ur Astrophysik Potsdam (AIP), Max-Planck-Institut f\"ur Astronomie (MPIA Heidelberg), Max-Planck-Institut f\"ur Astrophysik (MPA Garching), Max-Planck-Institut f\"ur Extraterrestrische Physik (MPE), National Astronomical Observatories of China, New Mexico State University, New York University, University of Notre Dame, Observat\'ario Nacional / MCTI, The Ohio State University, Pennsylvania State University, Shanghai Astronomical Observatory, United Kingdom Participation Group, Universidad Nacional Aut\'onoma de M\'exico, University of Arizona, University of Colorado Boulder, University of Oxford, University of Portsmouth, University of Utah, University of Virginia, University of Washington, University of Wisconsin, Vanderbilt University, and Yale University.

\section{Data availability}
The predicted 4XMM-DR9 catalogue is available in a repository and can be accessed using a unique identifier, part of it is shown in Table~10. It is available in paperdata at http://paperdata.china-vo.org, and can be accessed with http://paperdata.china-vo.org/zyx/table10.csv.

\end{document}